\documentclass[iop]{emulateapj}

\usepackage{amsmath}
\usepackage{natbib}
\usepackage{color}
\usepackage{hyperref}
\hypersetup{colorlinks=true,urlcolor=black,linkcolor=red,citecolor=blue}


\newcommand{\msun}{M_\odot}

\newcommand{\mpccm}{\ensuremath{m_\mathrm{p}\,\mathrm{cm}^{-3}}} 

\shorttitle{Feedback by AGN Jets and Winds}
\shortauthors{Dugan et al.}

\begin{document}

\title{Feedback by AGN Jets and Wide-Angle Winds on a Galactic Scale}

\author{Zachary Dugan\altaffilmark{1}, Volker Gaibler\altaffilmark{2}, Joseph Silk\altaffilmark{1,3,4}}
\altaffiltext{1}{The Johns Hopkins University Department of Physics \& Astronomy, Bloomberg Center for Physics and Astronomy, Room 366
3400 N. Charles Street, Baltimore, MD 21218, USA}

\altaffiltext{2}{Universit\"at Heidelberg, Zentrum f\"ur Astronomie, Institut f\"ur Theoretische Astrophysik, Albert-Ueberle-Str. 2, 69120 Heidelberg, Germany}
\altaffiltext{3}{Institut d'Astrophysique de Paris, UMR 7095, CNRS, UPMC Univ. Paris VI, 98 bis Boulevard Arago, 75014 Paris, France}
\altaffiltext{4}{Beecroft Institute for Cosmology and Particle Astrophysics, University of Oxford, Keble Road, Oxford OX1 3RH, UK}


\begin{abstract} 
To investigate the differences in mechanical feedback from radio-loud and radio-quiet Active Galactic Nuclei (AGN) on the host galaxy, we perform 3D AMR hydrodynamic simulations of wide angle, radio-quiet winds with different inclinations on a single, massive, gas-rich disk galaxy at a redshift of 2-3.  We compare our results to hydrodynamic simulations of the same galaxy but with a jet.  The jet has an inclination of 0$^\circ$ (perpendicular to the galactic plane), and the winds have inclinations of 0$^\circ$, 45$^\circ$, and 90$^\circ$.  We analyze the impact on the host's gas, star formation, and circum-galactic medium.  We find that jet feedback is energy-driven and wind feedback is momentum-driven.  In all the simulations, the jet or wind creates a cavity mostly devoid of dense gas in the nuclear region where star formation is then quenched, but we find strong positive feedback in all the simulations at radii greater than 3 kpc.  All four simulations have similar SFRs and stellar velocities with large radial and vertical components.  However, the wind at an inclination of 90$^\circ$ creates the highest density regions through ram pressure and generates the highest rates of star formation due to its ongoing strong interaction with the dense gas of the galactic plane.  With increased wind inclination, we find greater asymmetry in gas distribution and resulting star formation.  Our model generates an expanding ring of triggered star formation with typical velocity of order 1/3 of the circular velocity, superimposed on the older stellar population.  This should result in a potentially detectable blue asymmetry in stellar absorption features at kpc scales.

\end{abstract}

\section{Introduction}
Powerful Active Galactic Nuclei (AGN) have long been predicted and observed to have tremendous impacts on the galaxies that host them, from the scale of the galactic bulge to the Circum-Galactic Medium (CGM).  The well measured $M_\mathrm{BH}-\sigma_\mathrm{v}$ relationship is the first indicator of an important coevolution of the central black hole and the bulge, and several competing theories  aim to explain the tight correlation \citep{Silk98, Umemura01, Jahnke11}.  Jet-driven outflows can extend up to 50 kpc away from their source \citep{Shih15, Liu, Zakamaska13b, Nesvadba06}.  Within a single galaxy, the impact AGN feedback on the host's gas and star formation will vary depending on both the nature of the Interstellar Medium (ISM), how dense and clumpy the host's gas is, as well as on the type of feedback, radio-loud or quiet, etc. \citep{Kalf2012, Wagner13_out, Zinn13}.  The impact on star formation in particular has remained a mystery because of the difficult nature of its observation, especially at higher redshift.  Typically, negative feedback has been invoked to explain the lack of observed large, luminous galaxies predicted by $\Lambda$CDM theory \citep{Weinmann}.  However, both recent observations and simulations have begun to paint a more complex picture.

\subsection{Observations}
The observational approach to establishing a well defined relationship between AGN feedback and star formation can be difficult because the observational techniques employed to calculate star formation rates can be complicated by the presence of powerful AGN \citep{Zakamska16b}.  

Many observations have lead to the conclusion that AGN may quench star formation in the host.  \citet{Fabian12} describes how both radio-loud and radio quiet quasars drive bubbles and winds that simultaneously expel gas that might otherwise form stars from the galaxy and prevent the accretion of new gas onto the galaxy to form stars in the future.   This latter process can heat inter cluster gas and reduce star formation by an order of magnitude.  \citet{Schwamb16} provides evidence of negative feedback on the host through the expulsion of residual molecular gas.  \citet{Morganti15} shows that large amount of molecular gas can be driven by relativistic jets, although not always fast enough to be expelled from the galaxy. 

However, some observations indicate that AGN may not always act to quench star formation in the host.  \citet{Karouzos16} find evidence against AGN outflows as agents for negative feedback even at small redshifts, z $<$ 0.1, and low luminosities, L $< 10^{42}$ erg/s.  Using Gemini Multi-Object Spectograph data on six low redshift, type 2 AGN, they find that while the outflow velocities can reach 600 km s$^{-1}$, the $<$ 2.1 kpc size of the outflows are too small to quench an entire galaxy.   \citet{Labiano16} examines two low-redshift radio-loud AGN with outflows in different stages of the process.  Their calculated kinematics, star formation efficiency, and star formation rates indicate that AGN feedback is not necessarily responsible for the {apparently} low SFR in {evolved} AGN systems, but instead that perhaps the calculated SFRs are too low or that the estimated molecular gas content of these galaxies is too high.  

Other studies find a more complicated relationship between quasar winds and star formation in the host.  \citet{Carniani16} studies two quasars with fast outflows and observes star formation, but not in the path of the wind.  They conclude that the most likely possibility is one of simultaneous positive and negative feedback in the host, in which the outflows remove gas that could form stars along the direction of the wind while compressing gas around the edges of the outflow, triggering star formation.  They also postulate that several cycles of feedback could be necessary to quench star formation in the host completely.  Similarly, \citet{Cano12} observes AGN outflows quenching star formation along the path of the outflow and simultaneous star formation in the other parts of the galaxy.  

Some groups attempt to establish relationships between quasar outflow velocity and star formation rates and find mixed results.  \citet{Balmaverde16} observe 224 quasars at $z<1$ with outflows, and find that strong outflows have slightly higher SFRs than weak outflows at similar redshifts.  \citet{Wylezalek16}  also observe 133 radio-quiet quasars with outflows and use the [OIII]$\lambda5007\AA$ line width to determine outflow velocity.  They find a positive correlation between outflow velocity and star formation rate.  They also examine correlations between outflow velocity and specific star formation rate (sSFR), and observe no correlation for the overall sample but a negative correlation for those galaxies with SFRs > $100$ M$_\odot$ yr$^{-1}$.  They postulate that these galaxies have higher gas content because of the higher SFR, and that AGN feedback has more of a negative impact with respect to star formation in gas-rich galaxies.  Also important, the study shows a positive correlation between AGN luminosity and outflow velocity.  

Other studies find a lack of correlation between AGN luminosity and SFR.  \citet{Pitchford16} document 513 luminous type 1 quasars with extreme star formation rates and find that for a given redshift, the SFR does not vary with AGN luminosity, black hole mass, or Eddington Ratios.  They find that star formation in (HiBal's) is not impacted by outflows and conclude that for 0 < z < 3 star bursts in quasars typically evolve as the would without the presence of the AGN.  

Still other observations indicate that radio-loud quasars are more likely to trigger star formation than their radio quiet counterparts.  Analyzing almost 20,000 quasars from the Sloan Digital Sky Survey (SDSS), \citet{Kalf2012} use [OII] emission lines to estimate SFRs in quasars with and without jets.  After finding higher SFRs in the radio-loud AGN, they conclude that the jets trigger star formation. \citet{Zinn13} combine far-infrared and radio data on several hundred AGN from the Chandra Deep Field South to examine differences in star formation because of AGN jets.  Using the far-infrared data as a tracer for star formation, they find a correlation between enhanced SFRs and radio-loud quasars, even when compared to radio quiet quasars with similar luminosities.  Their results indicate positive feedback from the mechanical energy of jets and negative feedback from the photo-dissociation and heating of molecular gas.  

Some observers find evidence for AGN-triggered star formation on smaller scales, in Giant Molecular Clouds (GMCs) and smaller clouds alike.  \citet{Tremblay16} presents observations from ALMA showing AGN jets can act as mechanical pumps for giant molecular clouds, blowing them away with jet-driven bubbles before gravity pulls them back.  In these observations, the outer regions of the molecular clouds show star formation, possibly triggered by the expansion of the jet bubble.  \citet{Cresci15} use the Measuring Active Galactic Nuclei Under MUSE Microscope (MAGNUM) survey and present evidence for positive feedback from NGC 5643, a radio-quiet AGN with outflows.  They observe double sided ionization cones with high-velocity gas and star formation in clumps exposed to the resulting outflow, and they propose the compression of the clouds from the outflow is triggering the star formation.  The clouds are located at 1.2 kpc.  The projected velocity of the outflow is 423 km s$^{-1}$.  They also find a ring of star formation at 2.3 kpc, which agrees well with theoretical studies from \citet{Gaibler12} and \citet{Dugan14}.

\subsection{Theoretical Work}

\citet{Wagner16} reviewed theoretical work on both positive and negative feedback from radio-loud and radio-quiet AGN. They conclude that the result depends on the geometry and density of the ISM, that spherical distribution of clouds and lower density of clouds cause negative feedback, while disk configurations and higher density clouds are more conducive to positive feedback.  \citet{Ishibashi12} also provide a theoretical framework for AGN-triggered star formation, one in which stars are formed at increasingly large distances from the center of the galaxy, an ``inside out'' growth of star formation, and \citet{ZubovasKing16} reach a similar conclusion through analytic theory.  This process is also seen in the computational studies on radio-loud AGN simulations from \citet{Gaibler12} and \citet{Dugan14}.

To determine the feedback through pressure confinement of a jet-driven bubble, \citet{Bieri15} increased the pressure of the ambient gas around a disk galaxy to circumvent the computational challenges posed by the velocities and resulting shocks of an actual jet.  They calculate self gravity, and find the pressure causes an increased fragmentation of dense clouds in the host galaxy and subsequent increase in star formation (positive feedback).  \citet{Zubovas14_gal} provide analytic theory of galaxy-wide outflows and find that rapid cooling in the outflow leads to a two-phase gas and subsequent star formation.  

Some simulations focus on AGN feedback on smaller scales, such as shocks from jets or winds striking clouds.  \citet{Zubovas14_pressure} find that over-pressured shocks striking gas clouds causes fragmentation and star formation.  \citet{Dugan16} finds a threshold ram pressure beneath which over-pressured, high-velocity shocks with perturbations cause gas clouds to collapse and form stars.  

The simulations of AGN--cloud interaction by \citet{Dugan16} show that outflows can trigger star formation in gas clouds with significant radial velocity.  This work corroborates previous simulations from \citet{Gaibler12}, \citet{Dugan14}, and \citet{Zubovas13} and analytic theory by \citet{Silk12}.  Interestingly, \citet{Brown12} report that the orbits of many high-velocity stars (HVS) appear to emanate from the center of our own galaxy, which may agree with some HVS being caused by previous periods AGN of activity.

In this paper, we examine the role of opening angle and inclination in mechanical AGN feedback through four simulations, each on the same galaxy with different feedback parameters: a jet with a small opening angle and 0$^\circ$ inclination with respect to the disk, a wide angle wind with 0$^\circ$ inclination, a wide angle wind with a 45$^\circ$ inclination, and a wide angle wind with a 90$^\circ$ inclination.  We analyze the morphology and evolution of the galaxies, feedback to the host's gas, impact on star formation, and feedback to the circum-galactic medium (CGM).      

This paper is organized as follows.  In Section \ref{sec:analytical_exp_model}, we review analytic theory of bubble expansion.  We describe the simulation's numerics, set-up, and AGN feedback implementation in Section \ref{sec:simulations} and the analysis in Section \ref{sec:analysis}.  We show the results of simulations in Section \ref{sec:results}.  We present our Discussion in Section \ref{sec:discussion}, and conclude in Section \ref{sec:conclusion}.  

\section{Analytical Expansion Model} \label{sec:analytical_exp_model}
For a spherically symmetric galaxy, the energy conserving analytical models of a jet or outflow driven bubbles depend only on the power of the jet or outflow \citep{Bicknell96}.  The equations for the bubble's radius and resulting wind velocity are:
\begin{equation} \label{eq:bubble_radius}
R_\mathrm{b} = At^{3/5}
\end{equation}
\begin{equation} \label{eq:bubble_velocity}
v_\mathrm{w} = (3/5)At^{-2/5}
\end{equation}
where:
\begin{equation} 
A = \left(\frac{125P_\mathrm{jet}}{384\pi\rho_\mathrm{a}}\right)^{1/5}
\end{equation}

\begin{figure}
\includegraphics[width=1.\columnwidth,angle=0]{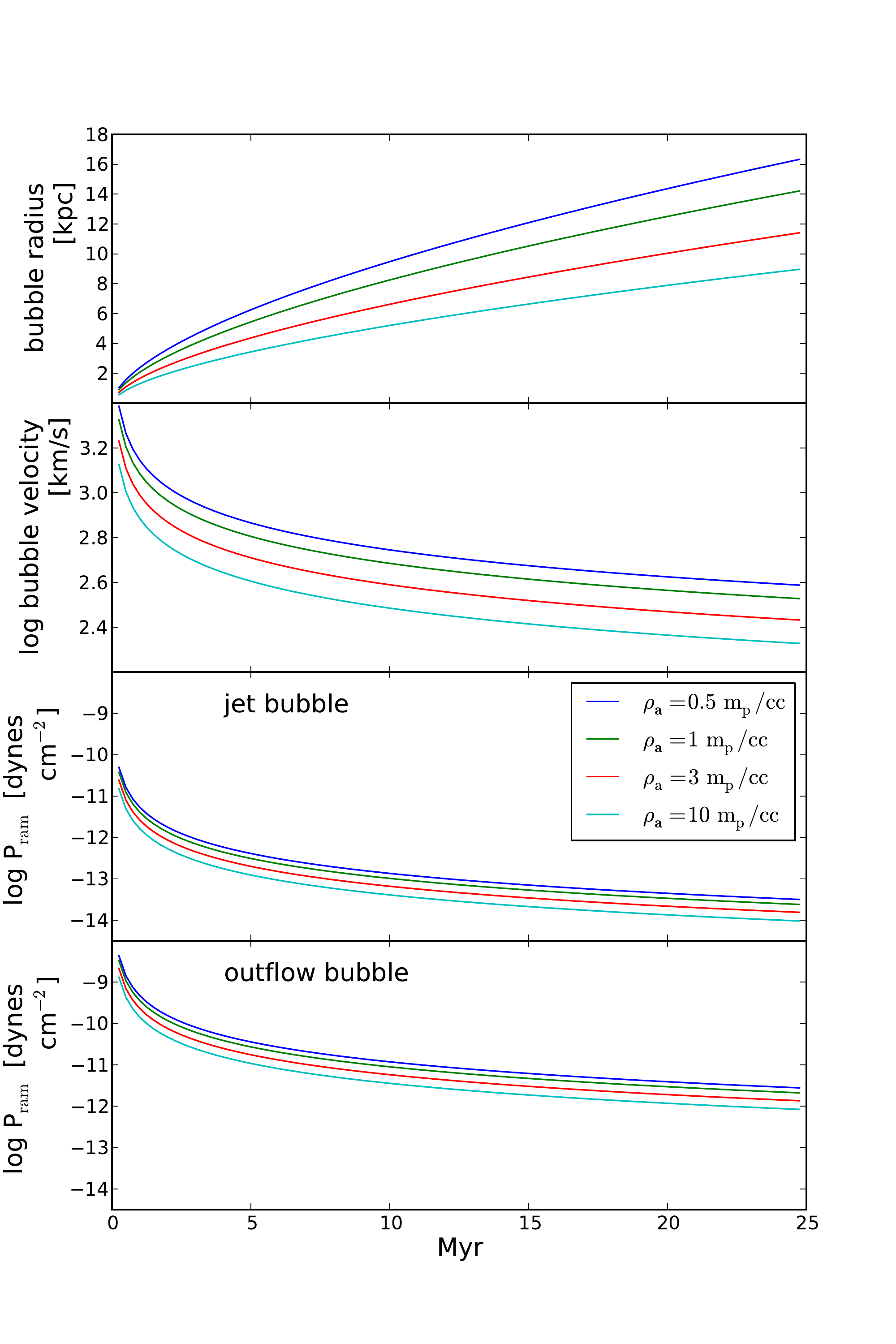} \\
\caption{Theoretical velocity, radius, and ram pressure of a jet or outflow driven bubble.  The bubble radius and velocity are calculated from Equations \ref{eq:bubble_radius} and \ref{eq:bubble_velocity} using a jet or outflow power of $5.5\times 10^{45}$ ergs s$^{-1}$.  The ram pressures are calculated using the mass outflow rates of a jet and outflow of the same power as in \citet{Gaibler12} and this study to calculate bubble density using Equation \ref{eq:bubble_density}.  }
\label{fig:rad_vel_ramp}
\end{figure}

If we assume that the bubble density is a function of the mass flux from a jet or outflow, and use Equation \ref{eq:bubble_radius} can be rearranged and integrated, as in \citet{Wagner12_jet}, to get the bubble density:
\begin{equation} \label{eq:bubble_density}
\rho_b = \frac{3}{4\pi}A^{-3}\dot{M}_\mathrm{jet}t^{-4/5}
\end{equation}
Now, combined with information on $\dot{M}_\mathrm{jet}$ and Equation \ref{eq:bubble_velocity}, the ram pressure from the bubble can be calculated.  If we assume a wind with the same power as the jet from \citet{Gaibler12}, $5.5\times10^{45}$ erg s$^{-1}$, the mass fluxes are 0.15 and 13.32 $M_\odot$ yr$^{-1}$ for the jet and wind, respectively.  We show the bubble radius and velocity for several ambient densities, as well as the ram pressure calculated from the mass fluxes listed above in Figure \ref{fig:rad_vel_ramp}.  

\section{Simulations} \label{sec:simulations}
Observations and simulations together capture how complicated and variable AGN feedback on star formation can be depending on redshift, the power of the AGN, the gas content of the host, whether the AGN is radio-loud or radio quiet, and many other factors.  In this study, we seek to refine understanding of the differences in mechanical feedback between jets and AGN winds, as well as the importance of AGN wind inclination on feedback.  We examine the impact of a jet and of AGN winds at three different inclinations have on the same disk galaxy's gas distribution and velocity, star formation rates and stellar velocities, as well as potential impact on the CGM.

\defcitealias{Gaibler12}{Paper I}
\subsection{Numerics and set-up}
We build on the four simulation runs from \citep*[][hereafter \citetalias{Gaibler12}]{Gaibler12} that examined AGN jet activity in a massive, gas-rich disk galaxy.  We construct the same thick, clumpy gaseous disk of $1.5\times10^{11}\msun$ with a scale radius $r_0 = 5$ kpc and a scale height $h_0 = 1.5$ kpc and hard cutoffs at $r = 16$ kpc and a height $h = 6$ kpc. The gas distribution, intended to mimic the clumpy interstellar medium, is derived from a fractal cube computed in Fourier space and the density profile

\begin{equation} 
\rho( \vec{x} ) \propto \exp \left\{ -r / r_0 \right\} \, \mathrm{sech}^2(h/h_0)
\end{equation}
for a log-normal probability distribution with an average density of 15.6 \mpccm and a median density of 10 \mpccm.  The Fourier power spectrum has a profile of $E(k)\propto k^{-5/3}$ for large wave numbers, greater than $h_0^{-1}$, to prevent large-scale asymmetries.  The ambient medium surrounding the galaxy has a density of 0.05 \mpccm.  The cooling function from \citet{SutherlandDopita1993} is computed for a metallicity of 0.5 $Z_\odot$ and employed with a temperature floor of $T/\mu$ = $10^4$ K, where $\mu$ is the mean particle mass in \mpccm.
Gravity was not included for the hydrodynamical runs due to the short time scales in the simulation compared to the disk evolution time scale and the lack of resolution on the very small scales, where collapse can occur on short time scales.

We utilize RAMSES \citep{Teyssier}, a second-order Godunov-type adaptive mesh refinement code.  The total computational domain is 128 kpc on a side with maximum resolution of grid cells of 62.5 pc on a side, refining wherever the cell to cell gradient exceeds 10\% in pressure or density (basically all regions of interest).  We employ the HLLC Riemann solver, the MonCen slope limiter, an adiabatic index of 5/3, and the "pressure fix" option, a hybrid approach that prevents negative pressures in regions with high Mach numbers.  

We employ the star formation model described by \citet{Rasera} which reproduces the Kennicutt--Schmidt relation \citep{Kennicutt1998}.  Stars are created only in regions where the number density of hydrogen is $n_\mathrm{H} > n_\star$, where $n_\star$ is the star formation threshold, with a rate controlled by a fixed star formation efficiency value and the local free-fall time:
\begin{equation}    
\dot{\rho}_\star =  \epsilon \rho / t_\mathrm{ff}  \,
\end{equation}    
where $\dot{\rho}_\star$ is the star formation rate, $\epsilon$ is the star formation efficiency, and $t_\mathrm{ff}$ is the local free-fall time.  Star particles are formed, and their mass is removed from the gas in the host cell.   Values of $\epsilon=0.05$ and $n_\star=5$ cm$^{-3}$ were chosen to yield the typical star formation rates of $\sim$ 150-200 $\msun$ yr$^{-1}$ without AGN activity.

\subsection{AGN Feedback}

In \citetalias{Gaibler12}, the bipolar jets were introduced by two adjacent cylindrical regions in the center of the disk that provide a collimated flow of gas with constant momentum input in both directions, a kinetic power of $5.5 \times 10^{45}$ erg s$^{-1}$, jet plasma density of $\rho_\mathrm{j}=5\times10^{-5}$ \mpccm, and jet velocity $v_\mathrm{j}$ = 0.8 c.  In this study, we perform three new simulations of wide-angle outflows from radio-quiet AGN in the same galaxy, with three orientations with respect to the disk.  These outflows have dual conical opening angles of 90$^\circ$, and inclinations to the disk plane normal of 0$^\circ$, 45$^\circ$, and 90$^\circ$, respectively.  In this paper, we label the simulations ``jet-i0'' for a jet with zero inclination to the disk, and ``wind-i0'', ``wind-i45'', and ``wind-i90'' for the wide-angle outflows with the three inclination values.  These outflows have the same power as the aforementioned jet, and radii of 1 kpc.  

When determining the parameters for these winds in our simulations, we start with the Eddington luminosity:
$L_\mathrm{Edd} = 4 \pi G M_\mathrm{BH} m_\mathrm{p} c / \sigma_\mathrm{T}$.  The mass of our black hole estimated from the dynamical masses of galaxies from \citet{Beifiori12}  is $\sim 1.5 \times 10^9 M_\odot$, which leads to $L_\mathrm{Edd} = 1.9\times 10^{47}$erg s$^{-1}$.  With the Eddington luminosity, we can determine the Eddington accretion rate $\dot{M}_\mathrm{Edd} = L_{edd}/(e c^2)= 4 \pi G M_\mathrm{BH} m_\mathrm{p} / (\sigma_\mathrm{T} e c)$ in which $e$ is the radiative efficiency.
Assuming $e$ of 10\%, $\dot{M}_{Edd} = 33.29$ $M_\odot$ $yr^{-1}$.  We assume that the ratio of mass outflow rate to the Eddington accretion rate is on the same order of magnitude as the Eddington ratio, as others in the literature have \citep{King}.   We employ an AGN outflow rate of 40\% of the Eddington accretion rate, a reasonable value for the Eddington ratio an AGN of this luminosity \citep{Shen08}, and we get an $\dot{M}_{out} = 13.32$ $M_\odot$ $yr^{-1}$.  To calculate the velocity and density of the outflow, we begin with the power of the outflow, which we set equal to that of the jet: $P_\mathrm{jet} = 5.5 \times 10^{45}$ erg s$^{-1} = 0.5 \dot{M}_\mathrm{out} v^2 $.  For an AGN outflow with the same power and the calculated mass outflow rate, we get the densities and velocities listed in Table \ref{tab:parameter_space}.  We set the pressure of the wind inside our conical injection region to the pressure of the jet, 1.6$\times10^{-10}$ dyne cm$^{-2}$, yielding nearly equivalent thermal powers of 2.4$\times10^{43}$ erg s$^{-1}$ for the jet and 4.1$\times10^{43}$ erg s$^{-1}$ for the winds.\\

\begin{deluxetable}{cccc} 
\tablecaption{Simulation Parameters}
\tablewidth{0pc} 
\tablecolumns{4}
\tablehead{ \colhead{Parameter}  & \colhead{Gaibler12} & \colhead{Wagner13} & \colhead{this study} }
\startdata
Resolution [pc] & 62.5 & 2 & 62.5\\  
v [km s$^{-1}$] & 240,000 & 30,000 & 36,203\\  
v [\%c] & 80 & 10 & 12.1 \\  
P [erg s$^{-1}$]& 5.5$\times10^{45}$ &10$^{44}$ & 5.5$\times10^{45}$ \\  
$\theta$ [$^\circ$] & $\sim 0$ & 30 & 45 \\  
$\dot{M}$ [M$_\odot$ yr$^{-1}$] & 0.15 & 0.1 & 13.32\\
$\rho_j$ [m$_\mathrm{p}$ cm$^{-3}$] & $5\times10^{-5}$ & 4.25 & $3.954\times10^{-3}$ \\    
r$_\mathrm{nozzle}$ [kpc] &0.4 & 0.01 & 1.0 
\enddata
\label{tab:parameter_space}
\end{deluxetable}

\section{Analysis} \label{sec:analysis}

To quantify feedback to the gas of the galaxy, we calculate several quantities.  We define the mechanical advantage as:

\begin{equation}
\mathit{MA}=\bold{p_\mathrm{r}}(t)/\int_0^t \bold{p_\mathrm{w}} dt=\bold{p_\mathrm{r}}(t)/(\dot{M}v_\mathrm{w}t)
\end{equation}
where $\bold{p_\mathrm{r}}(t)$ is the instantaneous radial momentum of the host's gas at time $t$, $\bold{p_\mathrm{w}}$ is the momentum of the wind, $\dot{M}$ is the wind's mass flux, and $v_\mathrm{w}$ is the wind's velocity.  This expresses the efficiency of the momentum transfer to the host's gas.  To quantify the manner in which the kinetic energy from the AGN is deposited to the host, we calculate the ratio of the kinetic to internal energy of the gas:
\begin{equation}
E_\mathrm{k}/E_\mathrm{int} = \rho v^2 (\gamma-1)/P 
\end{equation}
where $\gamma$ is the adiabatic index and $P$ is the pressure.  To the same end, we calculate the ratio of the kinetic energy of the cooler gas to the total energy injected into the galaxy:
\begin{equation}
E_\mathrm{k}/E_\mathrm{inj}=\rho v^2/(P_\mathrm{w} t)
\end{equation}
where $P_\mathrm{w}$ is the power of the wind.  We also estimate the velocity dispersion as a function of observation angle if calculated by an observer using absorption lines.  First, we establish a line of sight coming from the center of the galaxy to the observer.  We then take a line and oversample it by a factor of 3 more than the number of cells that would cover the line, extracting the density and velocity in the direction of the observer.  We then calculate the density-weighted velocity dispersion $\sigma_v$ as
\begin{equation} \label{eq:vel_disp} 
\sigma_v^2 = \cfrac{ \left( \left(\sum_i \rho_i v_{\mathrm{r}_i}^2\right)\left(\sum_i \rho_i\right) - \left(\sum_i \rho_i v_{\mathrm{r}_i}\right)^2    \right) }{ \left(\sum_i \rho_i\right)^2}
\end{equation}

\section{Results} \label{sec:results}
\subsection{Morphology and Evolution}
\begin{figure*}
\includegraphics[width=0.9\paperwidth,angle=0]{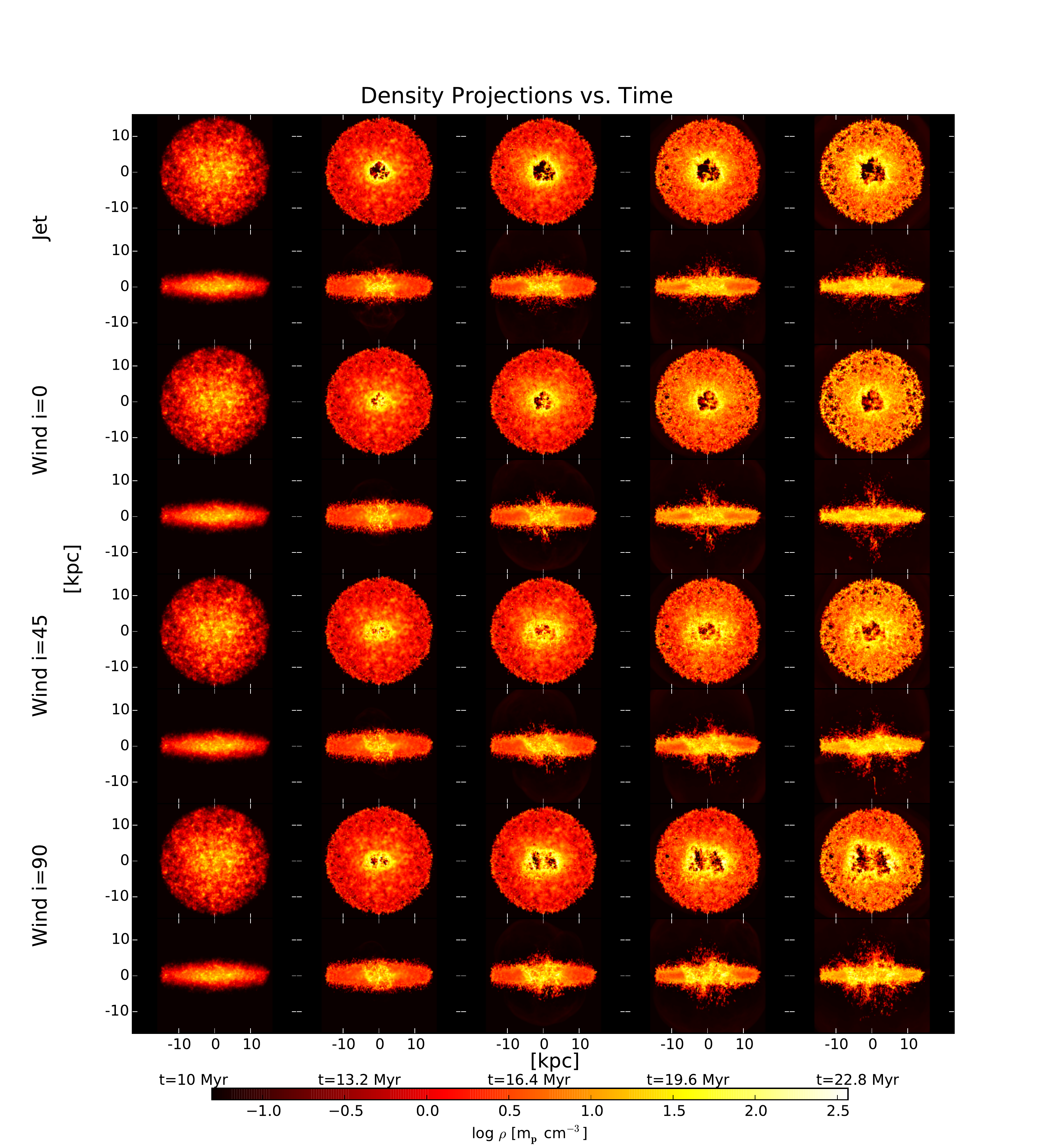} \\
\caption{Density projections.  Results from jet-i0 are in the top two panels, wind-i0 in the second two, wind-i45 in the third two, and wind-i90 in the fourth two.}
\label{fig:rho_projections}
\end{figure*}

In all of the simulations, the AGN feedback creates an extended cavity in the center of the galaxy.  Figure \ref{fig:rho_projections} shows face-on and edge-on projections of the density for all four simulations, and both the jet and the winds create rings of high density gas surrounding the cavity.  The regions where the winds directly strike the disk develop the highest densities, especially in the simulations where the outflow continuously interacts with the disk, wind-i45 and wind-i90.  The jet drills a far more diffuse hole in the center of the disk than do the outflows, which may result in a more distinct ring of star formation.  The outflows drive more mass off the disk, wind-i90 in particular.  This outflow creates an asymmetric cavity in the center of the galaxy as the wind continues to direct strong ram pressure out along the disk through the duration of the simulation.

\begin{figure*}
\includegraphics[width=0.75\paperwidth,angle=0]{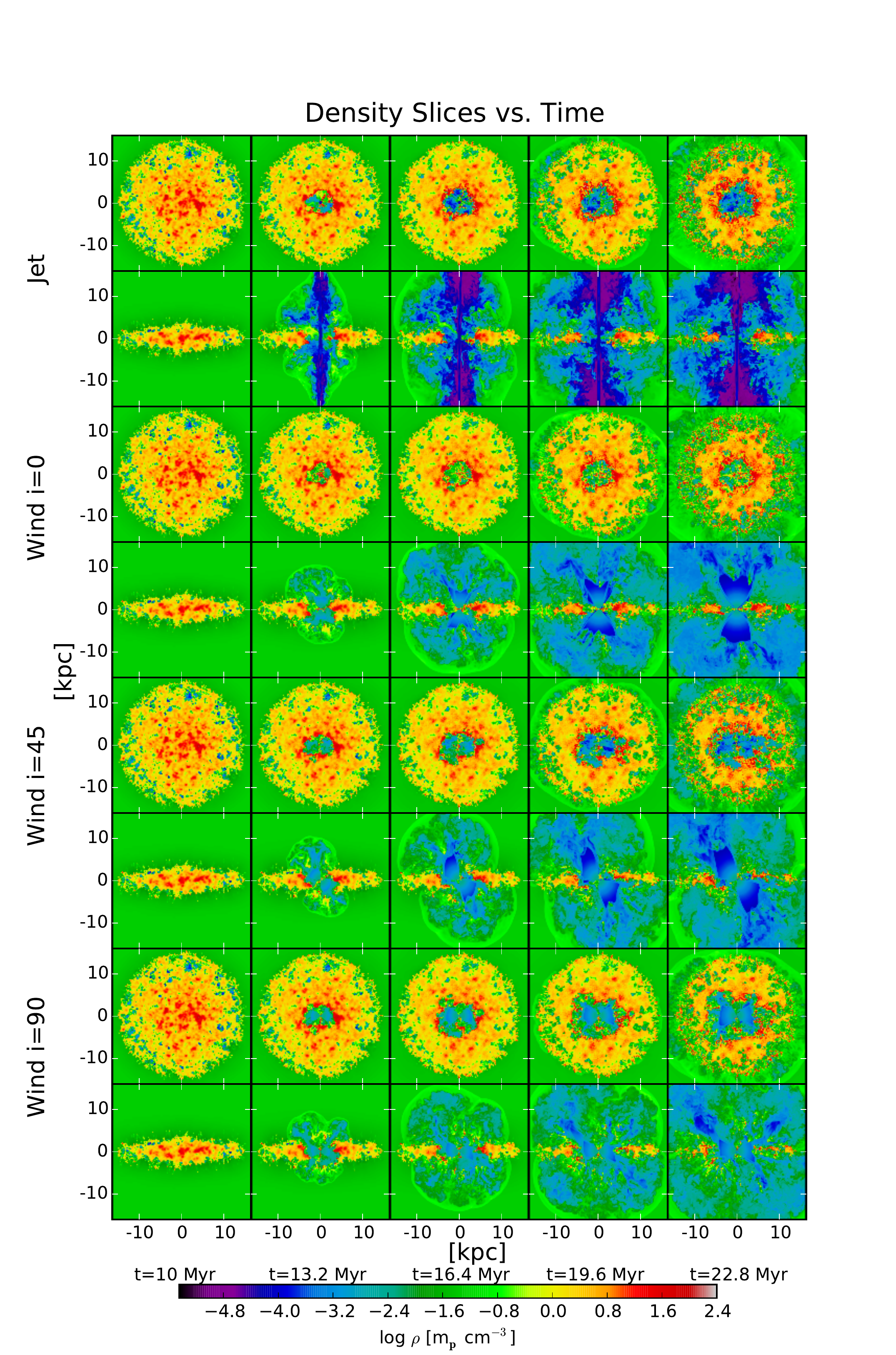} \\
\caption{Face on and edge on density slices through the center of the galaxy.  Results from jet-i0 are in the top two panels, wind-i0 in the second two, wind-i45 in the third two, and wind-i90 in the fourth two.}
\label{fig:rho_slice}
\end{figure*}

\begin{figure*}
\includegraphics[width=0.75\paperwidth,angle=0]{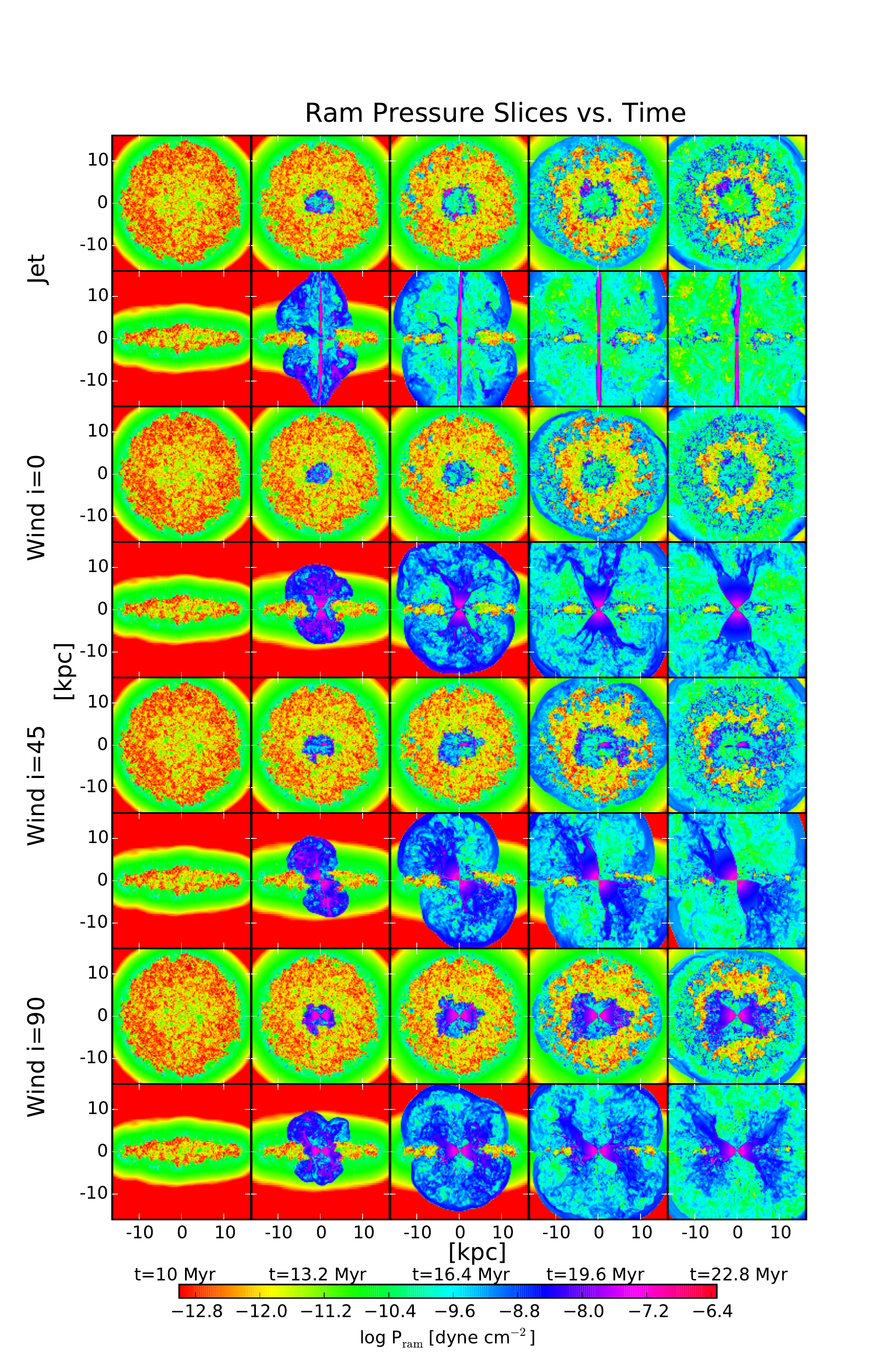} \\
\caption{Face on and edge on ram pressure slices through the center of the galaxy.  Results from jet-i0 are in the top two panels, wind-i0 in the second two, wind-i45 in the third two, and wind-i90 in the fourth two.}
\label{fig:ram_p_slice}
\end{figure*}

The winds break out of the disk between 2 and 2.25 Myr after initialization, whereas the jet breaks out after only 1.4 Myr, allowing its cocoon to begin growth and effect the host more quickly.  Only after the jet breaks through the disk, do the bubble evolution and feedback begin to differ from the winds.  The bubbles resulting from the winds seem to grow at a similar pace, as shown in Figure \ref{fig:rho_slice} which displays face on and edge on slices of the density for all four simulations.  The bubble growth from the winds is in agreement with the analytical theory of bubble expansion in a uniform environment, as shown in Figure \ref{fig:rad_vel_ramp}.  Though the jet itself has extreme ram pressure, the bubble it generates has a ram pressure typically an order of magnitude smaller than those generated from the winds, as shown in the ram pressure slices in Figure \ref{fig:ram_p_slice} and in agreement with analytical theory plotted in Figure \ref{fig:rad_vel_ramp}.  Figure \ref{fig:ram_p_slice} shows clearly that the high ram pressure in wind-i90 is directed straight into the high density disk for the duration of the simulation, continuously compressing this gas.  This makes a substantial difference in subsequent star formation, as we discuss in Section \ref{sec:feedback_to_stars}.

Although the jet breaks out of the disk more quickly, the cocoons in all the simulations have roughly the same velocity, $\sim$ 1,000 km s$^{-1}$.  The higher the inclination of the wind is, the more the clumpy ISM changes the direction and speed of the outflow.  In wind-i90 in particular, we see the development of asymmetric high velocity eddies and channels.  In jet-i0, the velocities of the dense gas along the disk are the smallest, even when compared to wind-i0.  In these two simulations, a thick ring of slow moving gas lies in the plane of the disk, with this ring being much thicker in the jet simulation.    
 
\begin{figure*}
\includegraphics[width=0.75\paperwidth,angle=0]{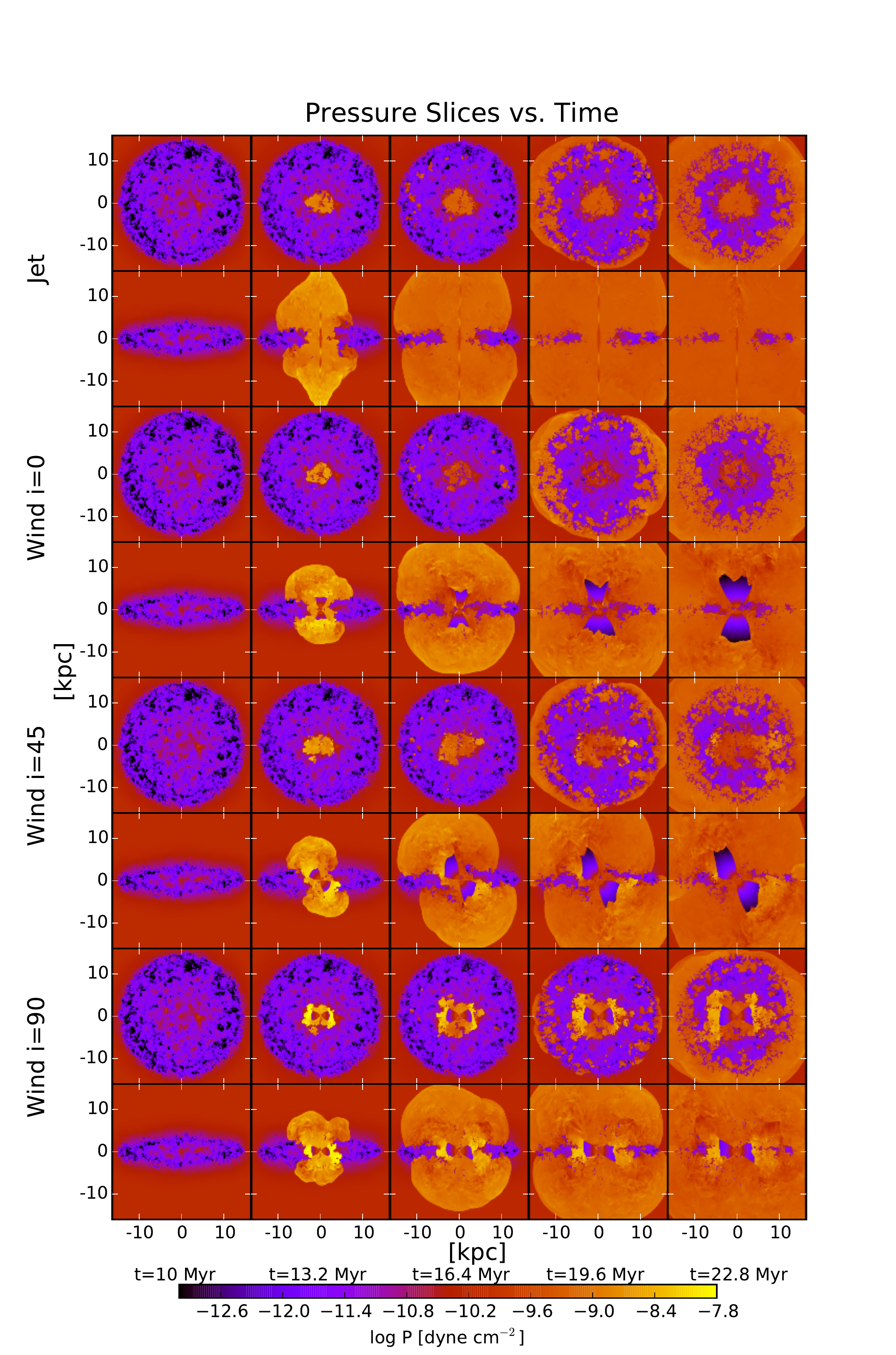} \\
\caption{Face on and edge on pressure slices through the center of the galaxy.  Results from jet-i0 are in the top two panels, wind-i0 in the second two, wind-i45 in the third two, and wind-i90 in the fourth two.}
\label{fig:pressure_slice}
\end{figure*}

\begin{figure}
\includegraphics[width=1.1\columnwidth,angle=0]{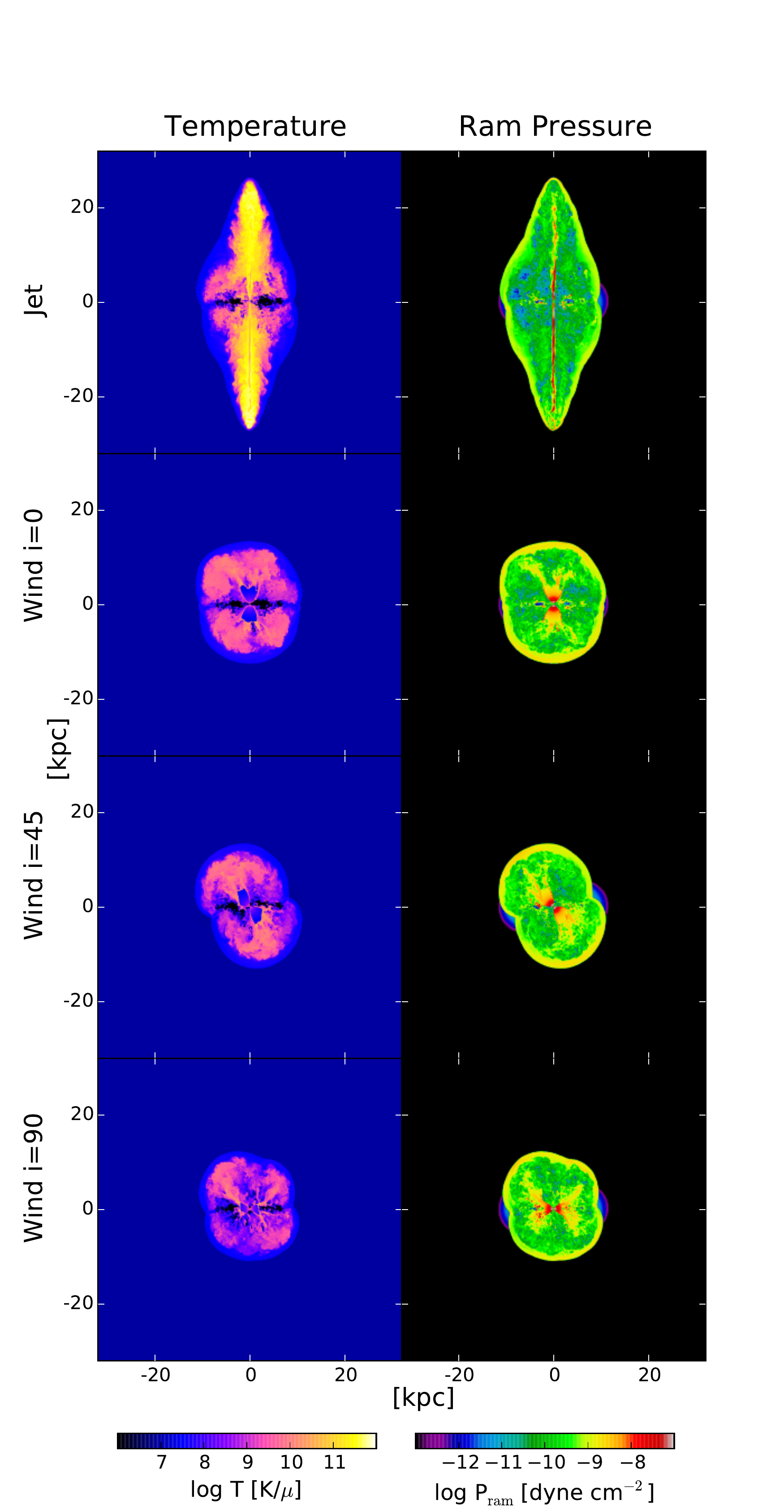} \\
\caption{Large scale temperature and ram pressure slices at 22.75 Myr.}
\label{fig:gal_temp_ram_p_slice_large2d}
\end{figure}

However, unlike the ram pressure, the jets initially produce regions of thermal pressure higher than the maximum pressure regions in the outflow simulations.  Figure \ref{fig:pressure_slice} shows face on and edge on pressure slices of all four simulations.  After about 10 Myr, the pressure within the jet cocoon and the outflow bubbles is about the same.  In wind-i0 and wind-i45, we see regions within the extended cone of outflow that are extremely under-pressured.  The outflow simulations with wind-i90, however, continues to create regions of high pressure where the wind strikes the disk directly.   The temperature within the wind bubbles are roughly the same, around $10^{10}$ K.  The jet, on the other hand, creates regions close to the jet but also close to the disk that are extremely hot, greater than $10^{11}$ K (but as expected for a hot plasma with highly relativistic electrons), shown in Figure \ref{fig:gal_temp_ram_p_slice_large2d}.  However, because the jet cocoon is under-dense relative to the wind cocoons, the thermal pressure is roughly the same.  We also see much greater fluctuation in temperature in the cocoon of wind-i90.

Our simulations show that jets are expected to have a greater impact on the CGM than winds for two reasons.  While winds create spherical bubbles, jet cocoons grow to a more conical shape with peaks that extend much further than the radius of the winds' bubbles.  The jet cocoon extends beyond 30 kpc and the wind bubbles achieve radii of only 16 kpc, roughly the radius of the galaxy.  Second, as a result of the low densities and high velocities within the jet cocoon relative to the wind bubbles, the jet cocoon has a much higher temperature.  Figure \ref{fig:gal_temp_ram_p_slice_large2d} shows slices of the temperature and ram pressure for all four simulations 12.5 Myr after the beginning of the AGN feedback.  Because of the extended structure and higher temperature, jets are more likely to impede accretion of gas onto the galaxy than winds are.  Additionally, the jet cocoon is more likely to strike an extra-galactic cloud or satellite because of its larger size.

\subsection{Feedback to the Gas}
We analyze AGN feedback to the host's gas in two density troughs, $\rho>0.1$ \mpccm and $\rho>1$ \mpccm, and annuli in the galaxy of 4 kpc radii.  Through the various quantifications of feedback to the gas in all four simulations, we see several common trends.  First, feedback to the more diffuse gas is typically stronger.  This means more efficient transfer of momentum and energy, and higher resulting velocities.  Second, as a result of the bubble's expansion through the disk in all the simulations, there is a time delay for feedback to the larger radii along the disk.  Third, the feedback from a jet happens much more quickly than in the simulations with the outflows. 

\begin{figure}
\includegraphics[width=1.\linewidth,angle=0]{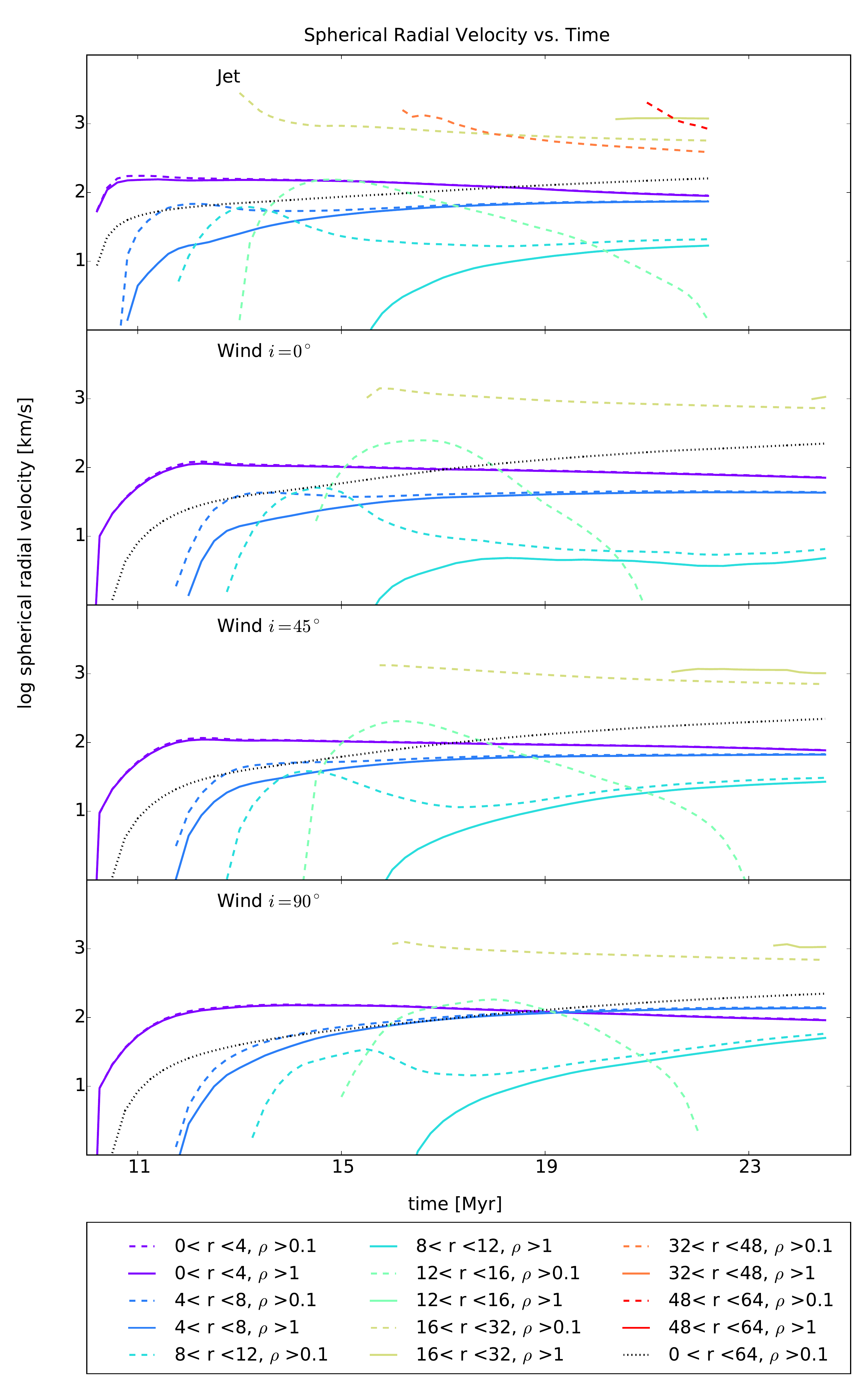} \\
\caption{Mass-weighted mean spherical radial velocity vs. time for various radial bins and densities.  Radii are in units of kpc, and densities are in units of \mpccm.  Results from jet-i0 are in the top panel, wind-i0 in the second, wind-i45 in the third, and wind-i90 in the fourth.}
\label{fig:spherical_velocity}
\end{figure}

The radial velocity of the host's gas in all the simulations reveals both more efficient feedback to the diffuse gas than the dense gas and the time delay to larger radii, as shown in Figure \ref{fig:spherical_velocity}.   For the dense gas at radii less than 8 kpc, the mass-weighted mean radial velocities reach velocities of 100 km s$^{-1}$ in all four simulations.  In all four simulations, the diffuse gas is accelerated to higher velocities at larger radii, typically reaching 100 km s$^{-1}$ between 8 and 12 kpc and 1,000 km s$^{-1}$ between 12 and 16 kpc.  These reflect the bubble itself, which comprises all the gas with densities greater than 0.1 \mpccm at radii greater than 16 kpc, essentially outside the original galaxy.  In fact, at radii greater than 16 kpc, the bow shock of the bubble can climb to densities greater than 1 \mpccm and has a velocity of 1,000 km $s^{-1}$, which impacts the CGM.  

Differences arise at larger radii.  The diffuse gas at radii greater than 16 kpc is accelerated to 1,000 km s$^{-1}$ by the jet in less than 4 Myr because of the large velocity of the jet and the bow shock of the jet.  In both jet-i0 and wind-i0, we see that the velocity of the dense gas at radii between 8 and 12 kpc ends up around 10 km s$^{-1}$.  However, as the inclination in the outflow simulations increases, the velocity of the same gas climbs to nearly 100 km s$^{-1}$ in wind-i90, indicating that the inclination of the outflow actually has a larger impact at larger radii.   

\begin{figure}
\includegraphics[width=1.\linewidth,angle=0]{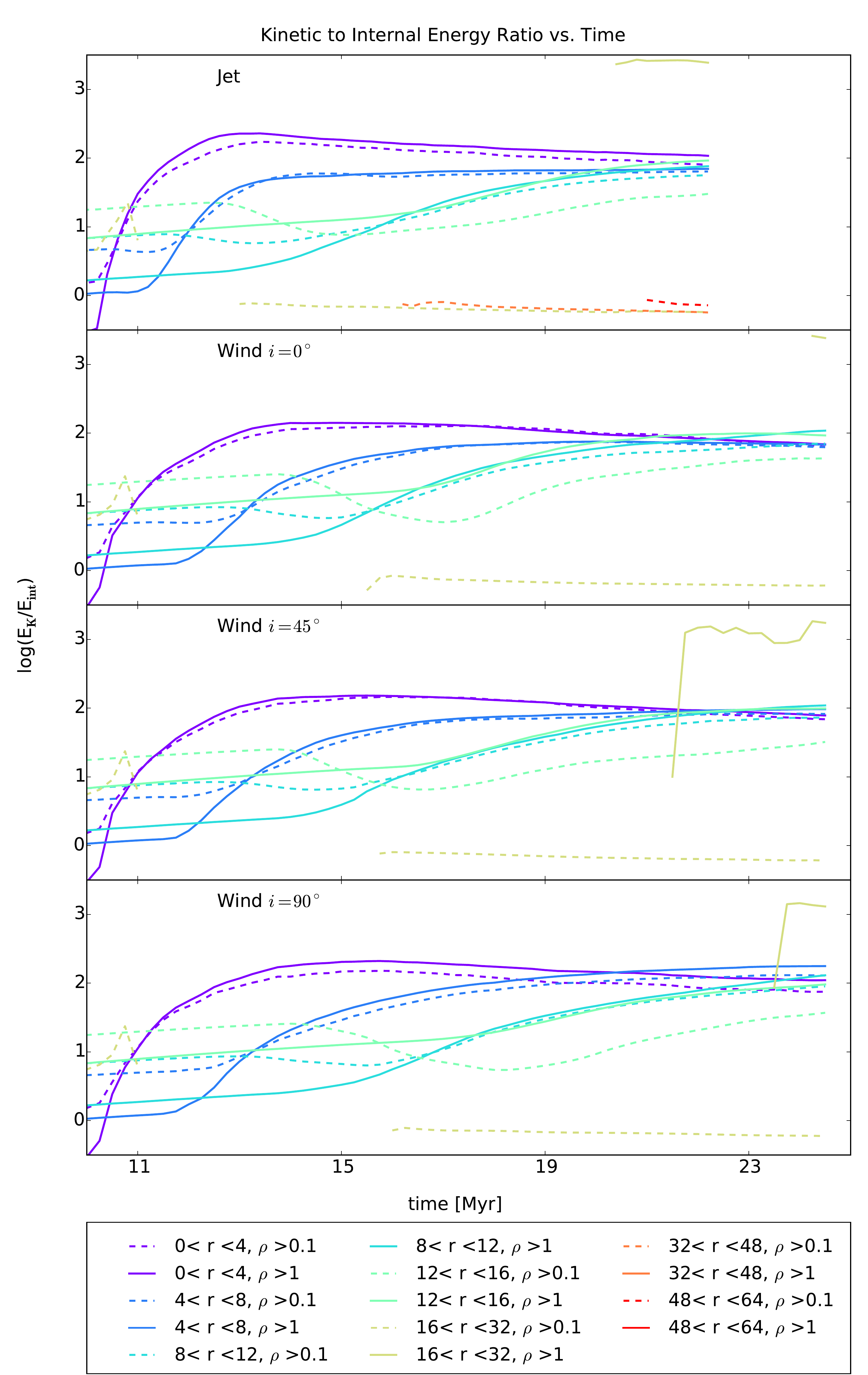} \\
\caption{Kinetic to internal energy ratio for various radial bins and densities.  Radii are in units of kpc, and densities are in units of \mpccm.  Results from jet-i0 are in the top panel, wind-i0 in the second, wind-i45 in the third, and wind-i90 in the fourth.}
\label{fig:ke_int}
\end{figure}

Most of the kinetic energy injected from the jet or wind into the host ends up as kinetic energy in the host's gas, rather than in its internal energy.  Figure \ref{fig:ke_int} shows the ratio between the kinetic energy and internal energy of the host's gas, a ratio that depends more strongly on radius than on density in all four cases.  Within 16 kpc, the ratio for both the dense and diffuse gas begins and ends with similar values, typically around a value of 100.  However, for the diffuse gas at radii larger than 16 kpc, more of the energy is thermal rather than kinetic, particularly in the case of the jet where high internal energy diffuse gas can be found at distances greater than 32 kpc from the galactic center.  Again, we see this ratio develops more quickly in the case of a jet than in the case of the winds, particularly at radii greater than 16 kpc.  

\begin{figure}
\includegraphics[width=1.\linewidth,angle=0]{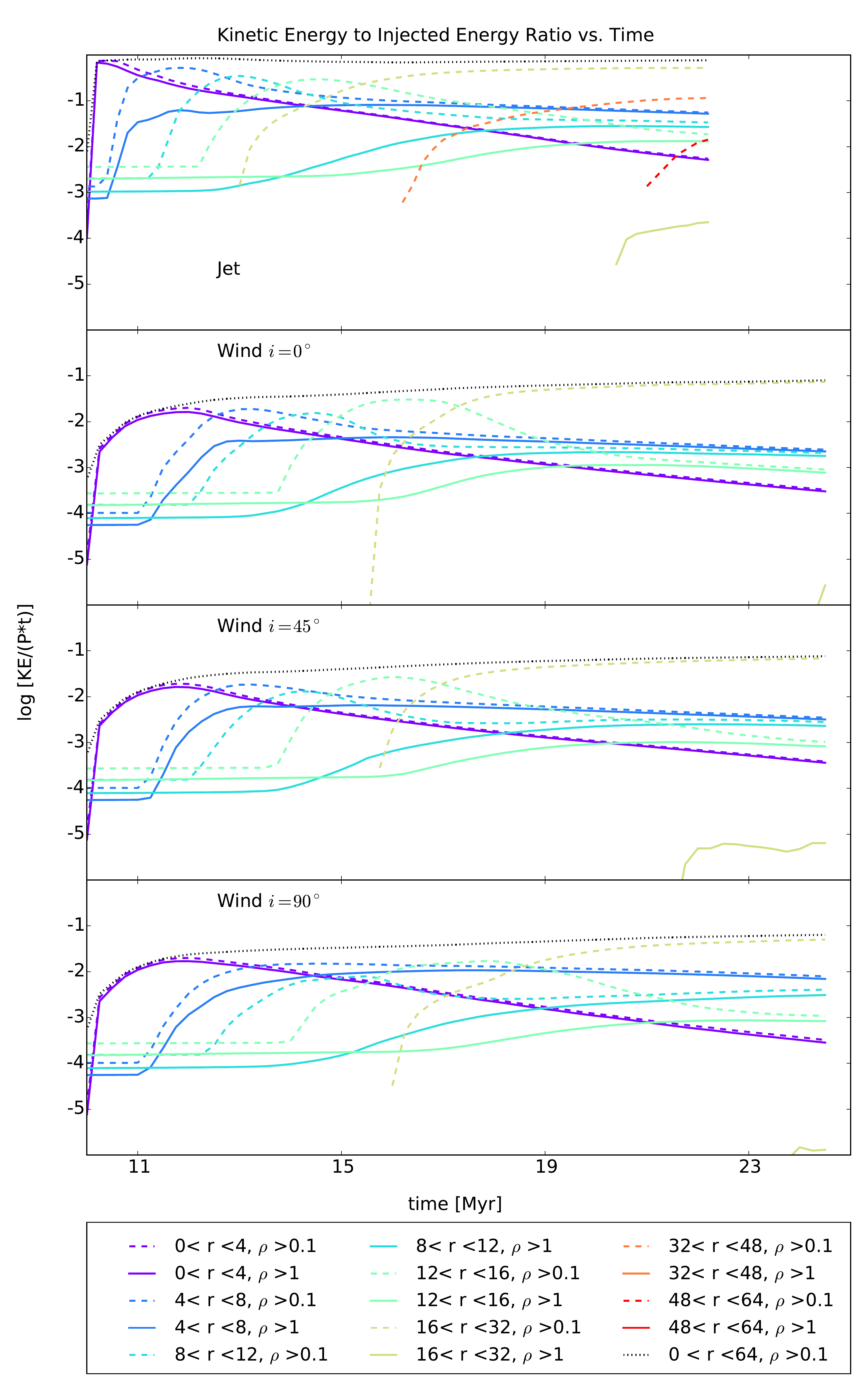} \\
\caption{Kinetic Energy to Injected Energy ratio vs. time for various radial bins and densities.  Radii are in units of kpc, and densities are in units of \mpccm.  Results from jet-i0 are in the top panel, wind-i0 in the second, wind-i45 in the third, and wind-i90 in the fourth.}
\label{fig:ke_pt}
\end{figure}

Most of the energy deposited by the jet or wind goes to the kinetic energy of the gas, and the efficiency with which kinetic energy and momentum are transferred to the host's gas shows that jet feedback is energy-driven and wind feedback is momentum-driven.  Figure \ref{fig:ke_pt} shows the ratio of kinetic energy of the host's gas to the injected kinetic energy.  Again, we see the jet and jet cocoon transfer kinetic energy to the host more quickly than the outflows, especially at the larger radii.  For all gas with a density greater than 0.1 \mpccm within a radius of 64 kpc, we see that this ratio approaches 1 for the jet, indicating the feedback is energy-driven.  However, for the same gas in the case of all of the winds, the ratio approaches 0.1, indicating a substantially less efficient energy transfer.  

\begin{figure}
\includegraphics[width=1.\linewidth,angle=0]{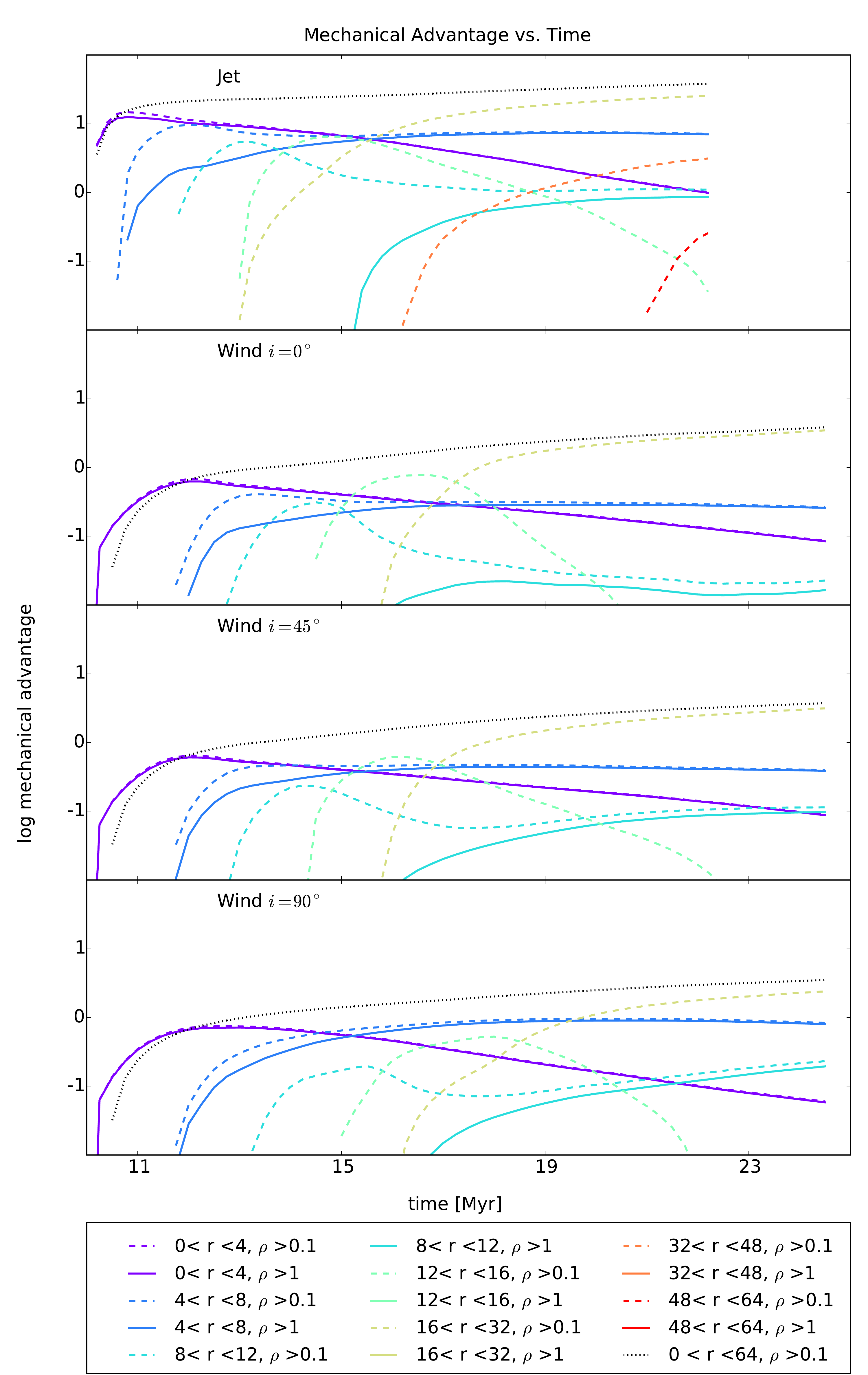} \\
\caption{Mechanical advantage vs. time for various radial bins and densities.  Radii are in units of kpc, and densities are in units of \mpccm.  Results from jet-i0 are in the top panel, wind-i0 in the second, wind-i45 in the third, and wind-i90 in the fourth.}
\label{fig:ma}
\end{figure}

The mechanical advantage, which is the ratio of radial momentum to the time-integrated injected momentum, reflects that the winds are also closer to momentum-driven.  Figure \ref{fig:ma} shows that for all gas with a density greater than 0.1 \mpccm within a radius of 64 kpc, the mechanical advantages of the winds end far closer to 1 than that of the jet.  Again, we see the jet transfers momentum to the host's gas faster than the winds do.  We see the mechanical advantage decrease with increasing radius for the dense gas and vice versa for the diffuse gas, which is not generally true for the kinetic energy to injected kinetic energy ratio.  

\begin{figure*}
\includegraphics[width=0.9\paperwidth,angle=0]{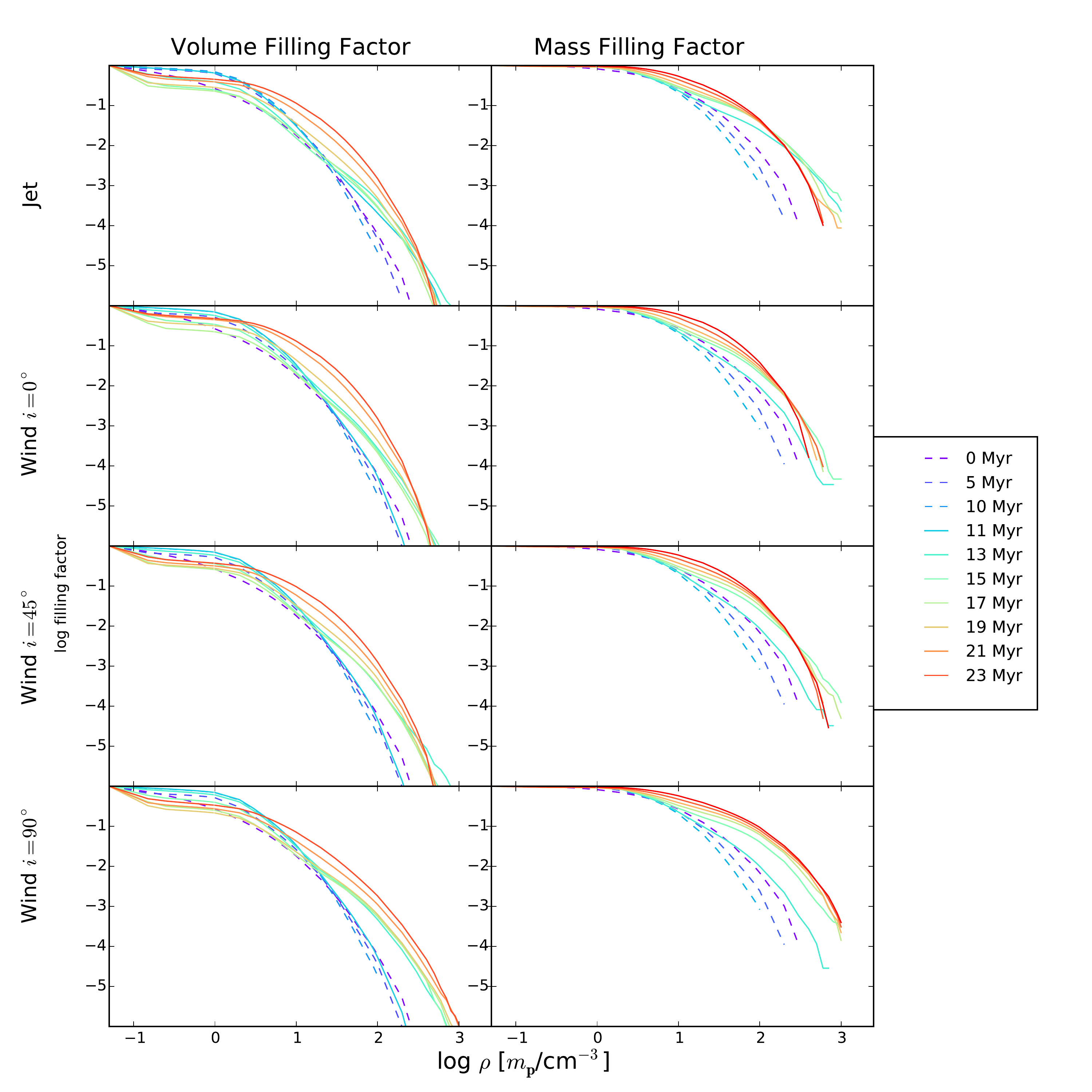} \\
\caption{Volume filling factor and mass filling factor for various times.  Dotted lines indicate times before the jet or wind has been initialized.  Results from jet-i0 are in the top panel, wind-i0 in the second, wind-i45 in the third, and wind-i90 in the fourth.}
\label{fig:filling_factor}
\end{figure*}

A helpful way of examining AGN impact on host morphology is to analyze the fraction of space and mass occupied by gas of certain densities.  The volume filling factor and the mass filling factor plots shown in Figure \ref{fig:filling_factor} show similar evolutions in which the volume and mass occupied by high density gas increase for 10-12 Myr after which they decrease, with the exception of wind-i90.  In that simulation, the volume and mass fraction occupied by high density gas continue to increase through the duration of the simulation.  The continuous increase is a direct result of the wind blasting directly into the disk, pushing the dense gas into more dense gas, which has important implications for star formation.   

\subsection{Star Formation} \label{sec:feedback_to_stars}

\begin{figure}
\begin{center}
\includegraphics[width=1.1\columnwidth,angle=0]{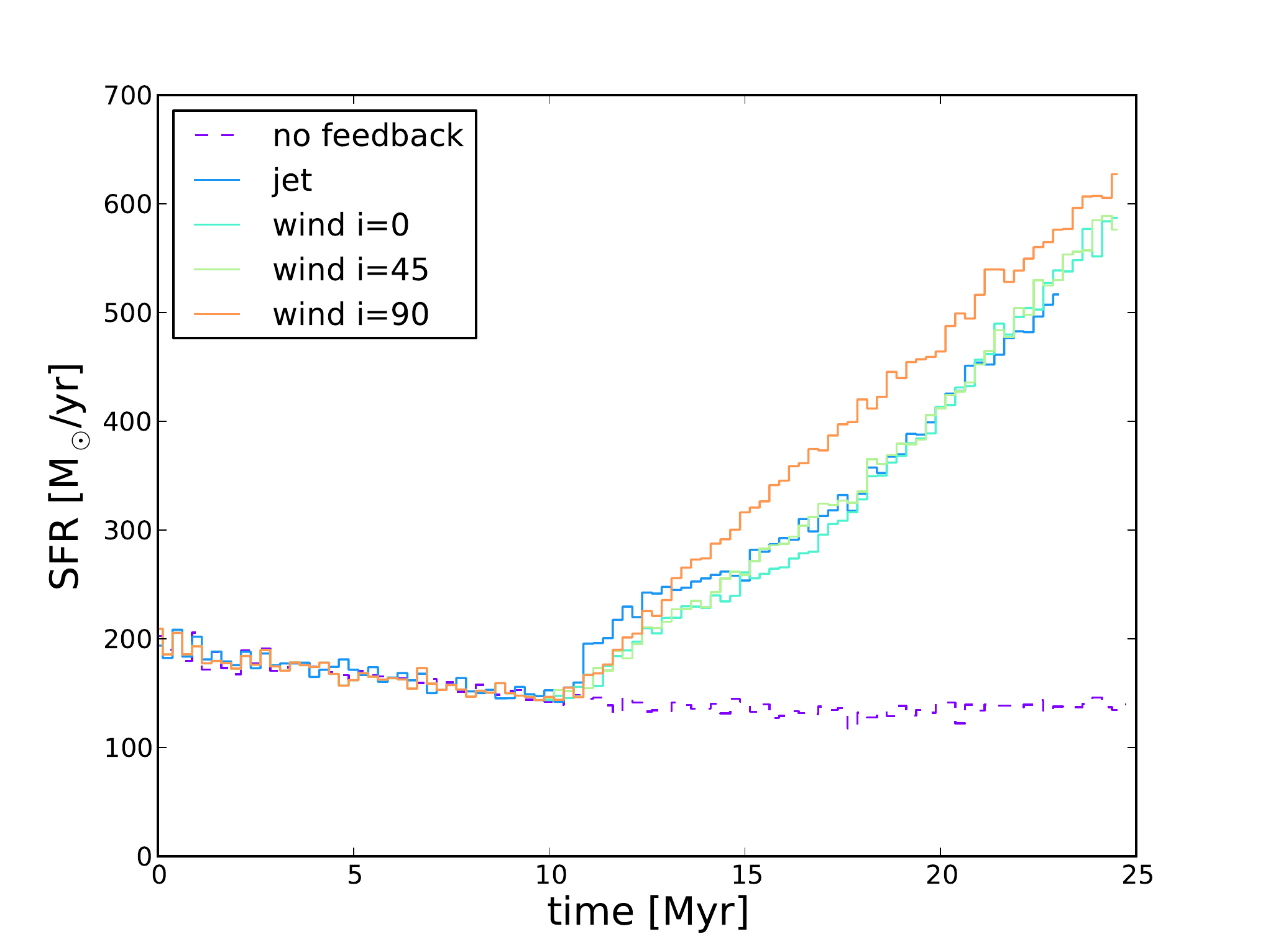} \\
\end{center}
\caption{Star formation rate.  Jet and winds initialized at 10 Myr.}
\label{fig:sfr}
\end{figure}

We find similar star formation rates for all the simulations except for wind-i90, which has an increased SFR as shown in Figure \ref{fig:sfr}.  Over the first 4 Myr, the SFR in the jet simulation outpaces the SFR in the wind simulations, which are all similar.  During this time the jet has not yet broken out of the galaxy, and thus all the kinetic energy goes toward very dense gas, creating regions of high density being compressed by high ram pressure and high thermal pressure, causing subsequent star formation.  However, after the jet breaks out of the galaxy, the rate of increase of SFR slows.  Around this time in the wind simulations, the winds break through this disk.  In wind-i90, however, the wind continuously pushes straight out along the disk and compresses the surrounding dense gas throughout the simulation.  The kinetic energy in the other simulations is no longer deposited directly into the disk, rather it goes into the bubble.  For this reason, only the SFR in the wind-i90 simulation  increases at the same rate throughout the simulation.  After 5 Myr, the increase of the SFRs in jet-i0, wind-i0 and wind-i45 slows down, but remain remarkably similar to one another.  With respect to star formation, only a significantly different inclination makes a difference, while the power of the wind or jet is the critical component.    

\begin{figure}
\includegraphics[width=1.1\columnwidth,angle=0]{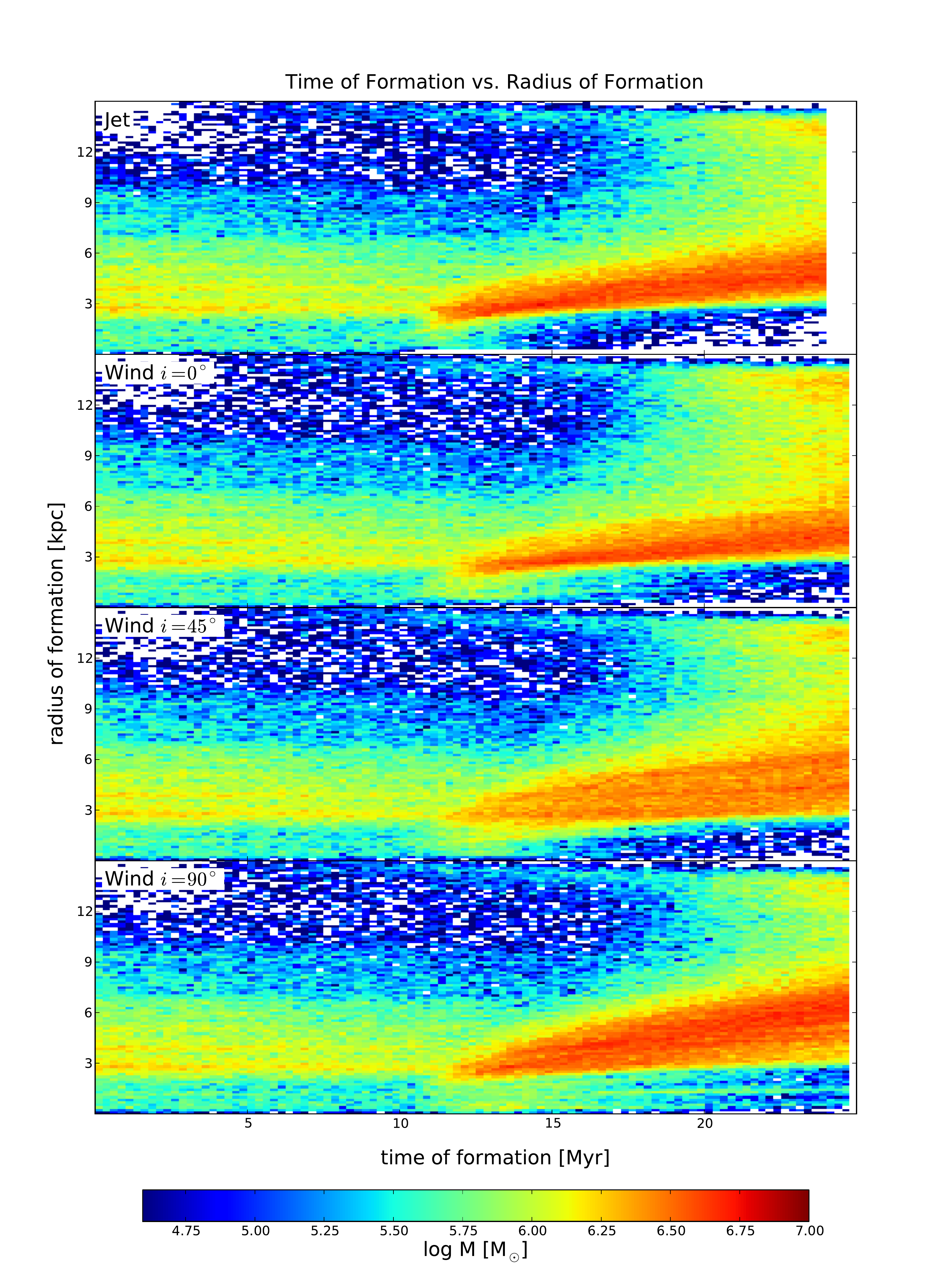} \\
\caption{Mass-weighted phase plot of radius of formation vs time of formation for star particles.}
\label{fig:time_disk_r}
\end{figure}

Analogous to the similarity of the SFR's in the four cases, the locations of stars formed during the AGN feedback are also similar.  Figure \ref{fig:time_disk_r} shows the radius of star formation versus the time of formation.  As in \citet{Dugan14}, a ring of star formation begins at a radius of roughly 2 kpc at 1 Myr after feedback begins and moves outward radially in all four simulations as a consequence of the bow shock from the cocoons expanding through the disk.  Also common to all four simulations is the stimulation of star formation at radii greater than 6 kpc beginning around 6 Myr after feedback begins resulting from the compression of the disk from the expanding bubble.  Not surprisingly, the pattern is extraordinarily similar in jet-i0, wind-i0, and wind-i45, with the final ring of star formation finishing between radii of 3 and 6 kpc about 15 Myr after the jet or wind initialization.  The main difference is again with wind-i90, where the wind pushes straight into the disk for 15 Myr.  In this simulation, the ring of star formation is closer to an oval extending from 3 kpc to nearly 9 kpc.  

\begin{figure*}
\includegraphics[width=0.8\paperwidth,angle=0]{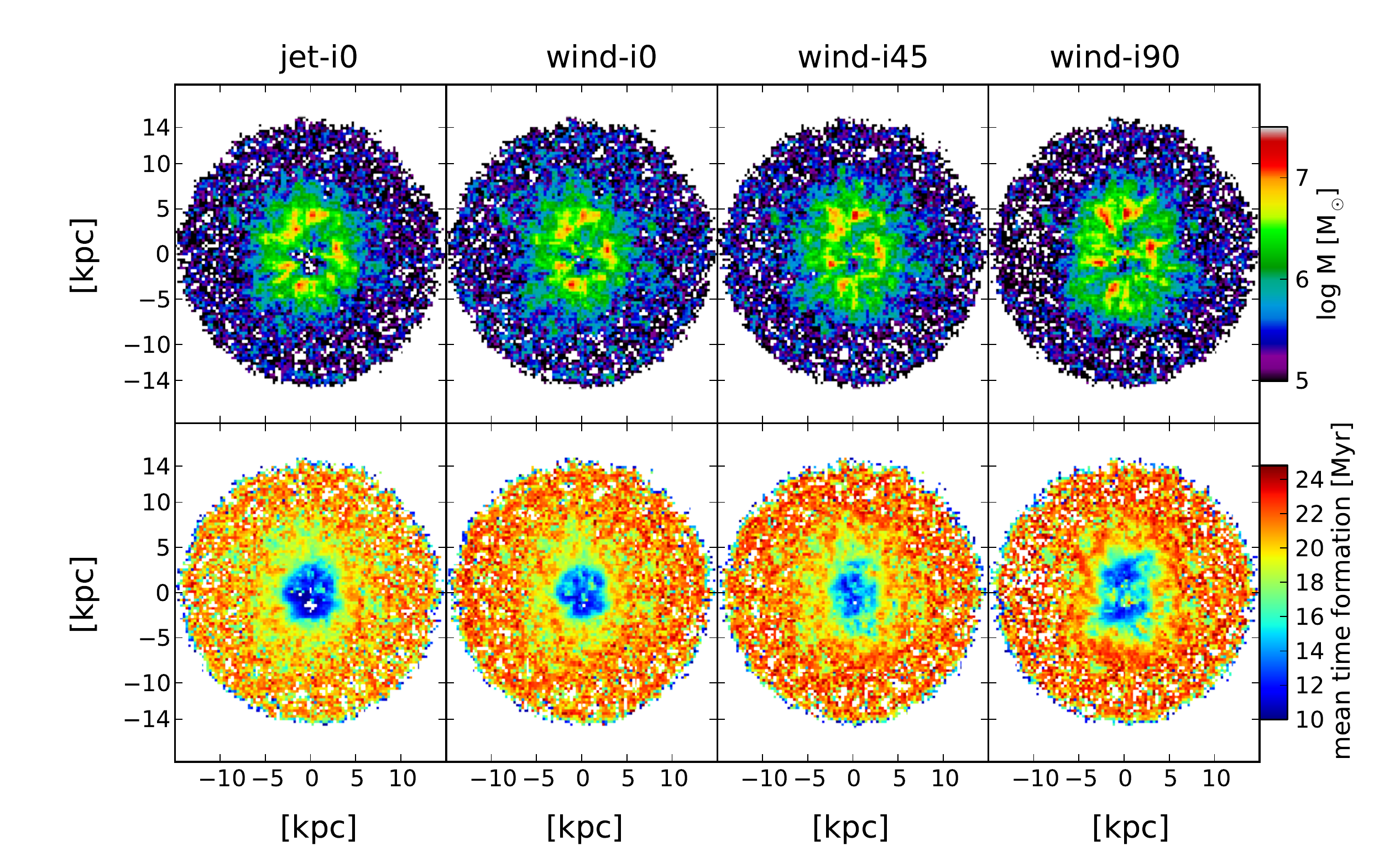} \\
\caption{Stellar mass distribution at time of formation, and mean time of formation for all star particles formed after AGN feedback is initialized at 10 Myr.}
\label{fig:star_time_mass}
\end{figure*}

Figure \ref{fig:star_time_mass} shows the mass-weighted locations of star formation along with the mean times of star formation for those locations.  It shows an inside-out pattern of star formation shown in Figure \ref{fig:time_disk_r} and discussed above.  In this central region, star formation is quenched quickly after the jet or wind begins in all four simulations, as reflected by the mean time of star formation.  The spatial distribution of star formation matches the spatial distribution of gas, including the asymmetries in wind-i45 and wind-i90.  In both jet-i0 and wind-i0, the jet and wind form a very circular cavity of gas in the central region of the galaxy, with star formation tracing the edge of this cavity, forming a clear circle.  In wind-i45, this region is more oval than in jet-i0 and wind-i0, and in wind-i90 it looks more oval still, tracing the direction of the wind.  Additionally, in wind-i90 more than any of the other simulations, the wind seems to have caused star to form along paths moving radially away from the galactic center, probably through accelerating clouds of gas that continue to form stars as they move.  This is because a higher fraction of the wind's kinetic energy is directed straight into the disk, compressing and accelerating clouds of gas that form stars.  

\begin{figure}
\includegraphics[width=1.1\columnwidth,angle=0]{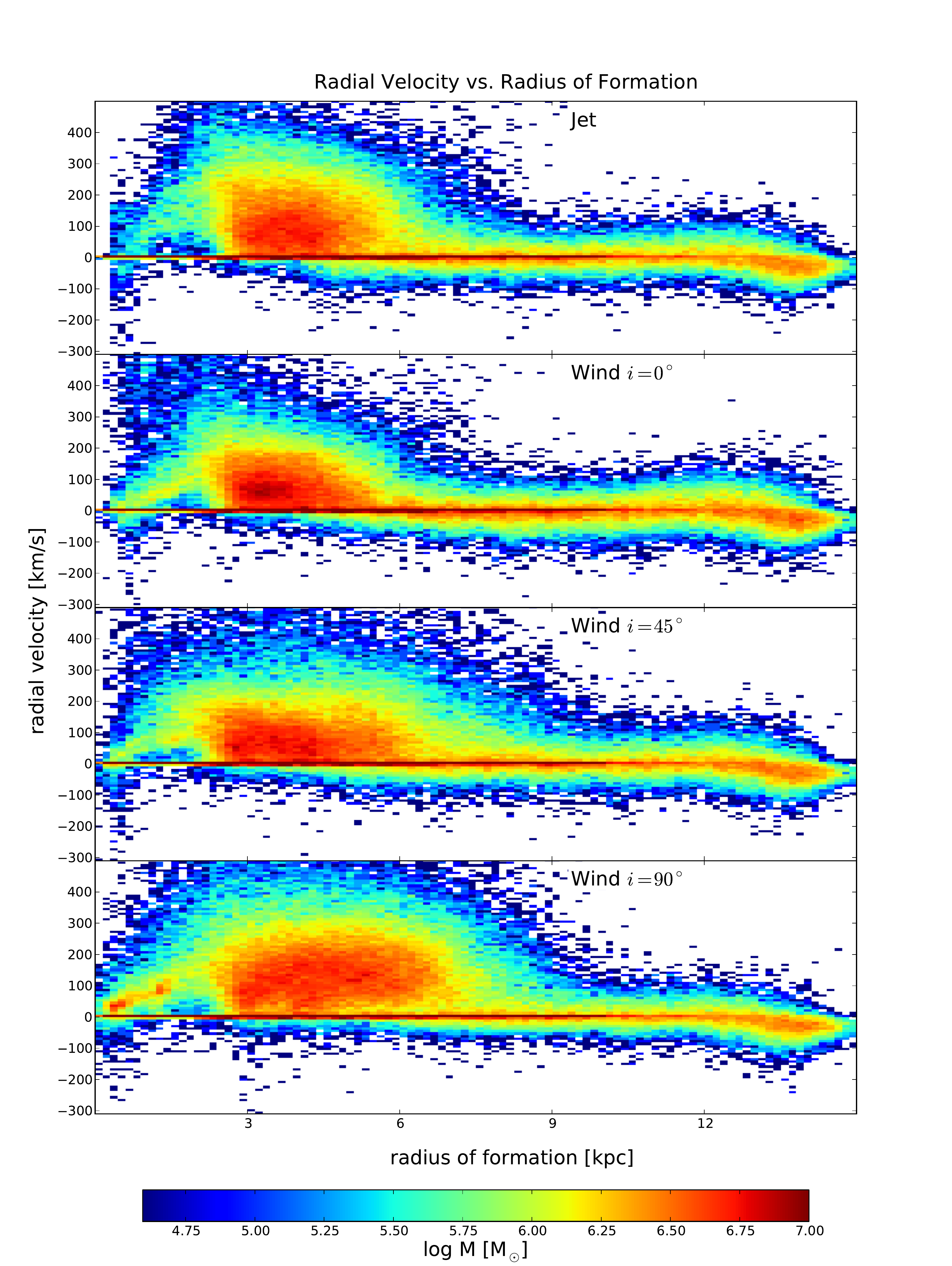} \\
\caption{Mass-weighted phase plot of radial velocity vs. radius for star particles.}
\label{fig:radial_v_r}
\end{figure}

Stars formed as a result of AGN feedback have positive radial velocities, particularly those formed in the dense rings of stars at radii from 3-6 kpc.  Figure \ref{fig:radial_v_r} is a phase plot of the radial velocity of formed stars versus the radius of formation.  The distribution of these velocities is similar between all the simulations, with the distribution of high velocities from wind-i90 extending further than in the other simulations, as reflected from the thicker ring of star formation shown in \ref{fig:time_disk_r}.  When analyzing this plot, it is important to remember that the star particles formed in these simulations are formed with the velocities of the gas in their birth cells, and it is reasonable to assume that the velocities of stars formed in these cells will be smaller.  Analysis in \citet{Dugan16} addresses star formation as a consequence of high velocity AGN winds and shows that the velocities of the resulting stars will not be as high as the wind or the surrounding gas.  

\subsection{Observing Velocity Dispersion}
\begin{figure*}
\includegraphics[width=0.9\paperwidth,angle=0]{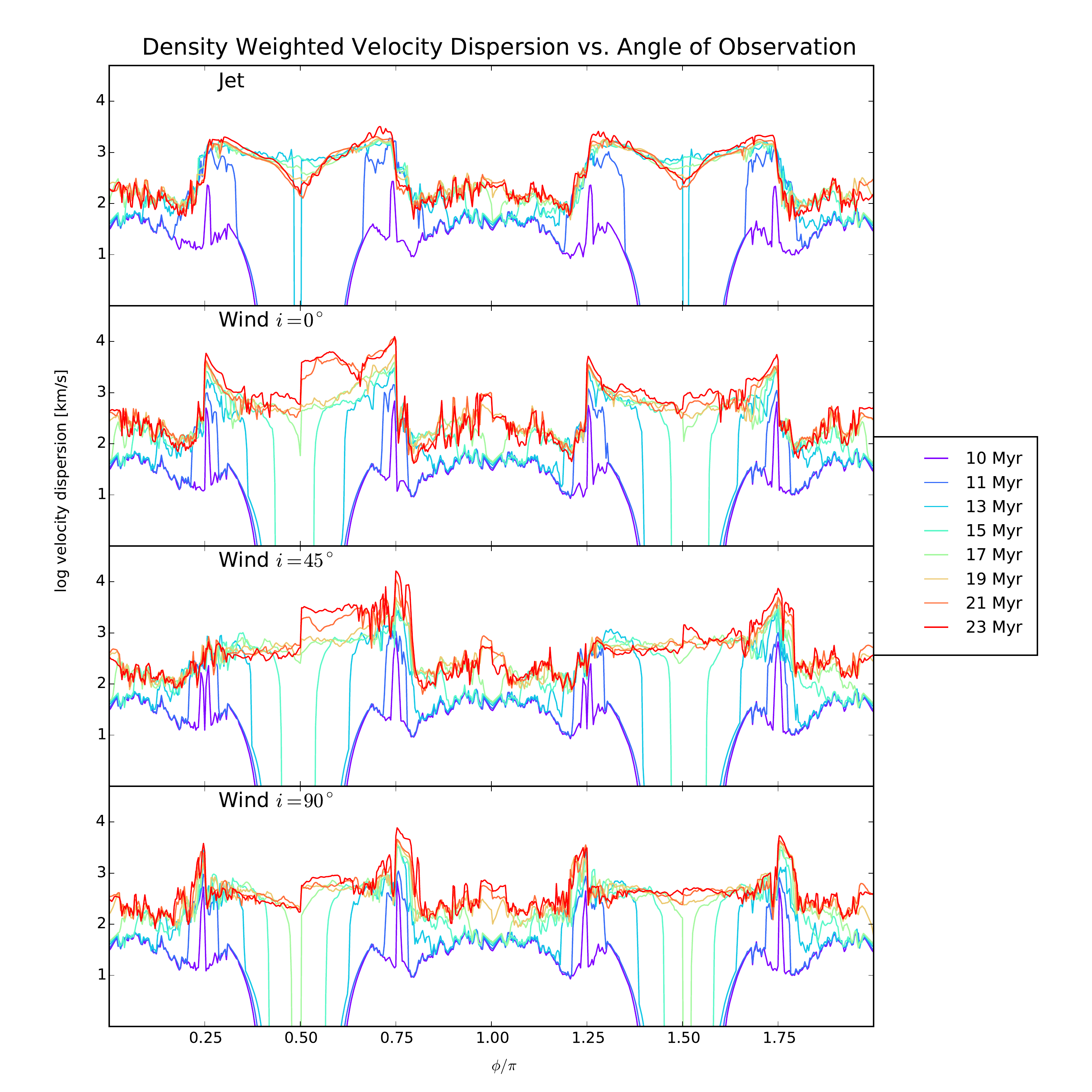} \\
\caption{Density-weighted Velocity Dispersion calculated from absorption lines.}
\label{fig:vel_phi_2}
\end{figure*}

Observations of AGN feedback are very difficult for a number of reasons, not the least of which is that some of the most interesting cases are at high redshift.  One of the better ways to quantify the internal dynamics of a host galaxy is to measure the velocity dispersion along the line of sight.  In theory, however, this approach can be problematic because of the angle of observation dependence and the asymmetric three-dimensional geometry of AGN.  To quantify this dependence on geometry, we show the density-weighted velocity dispersion of absorption lines as a function of observation angle for various times for all four simulations in Figure \ref{fig:vel_phi_2}.  The first conclusion is the clear dependence on angle of observation, particularly at earlier times.  Not surprisingly, both the jet and the wind-i0 simulations show much higher dispersions when looking down the jet or wind, a half angle of roughly 45$^{\circ}$, rather than looking at the galaxy edge on.  

In the first few Myr of feedback from both the jet and the winds, the observed velocity dispersion can vary 2.5 orders of magnitude depending on the angle of observation.  Between 1 and 2 Myr after the jet is initialized, velocity dispersions in the host can reach over 1,000 km s$^{-1}$ because the jet breaks through the disk in this window.  After 5 Myr, the maximum dispersion in the galaxy with the jet can exceed the minimum dispersion by up to 1.5 orders of magnitude, whereas in the case of the winds the disparity exceeds 2 orders of magnitude.  These results indicate that the angle of observation of absorption lines is important to the measured dispersions, particularly in the early stages.   

Another notable feature of this analysis is the temporal difference in the evolution of the velocity dispersion as a function of observation angle.  In the galaxy that hosts the jet, the velocity dispersion reaches its final distribution far more quickly than in galaxies with conical winds.  With the jet, this evolution takes just over 3 Myr after initialization, whereas with the winds it takes roughly 7 Myr.  This is a common theme with respect to feedback in general.  Because the jet breaks out of the disk faster than the winds, it deposits its kinetic energy and momentum to the host in a much shorter time than winds, as shown in Figures \ref{fig:ke_pt} and \ref{fig:ma}.

\section{Discussion} \label{sec:discussion}
Because the kinetic powers are all the same and the thermal powers are nearly the same, the jet and winds' interaction with the disk and resulting bubbles make the difference to the feedback.  We observe more thermalization in the case of the jet, leading to more energy-driven feedback.  In the case of the winds, the smaller disparity in densities results in less thermalization.  The feedback from the winds is momentum-driven and increasingly energy-driven with time, though not nearly to the same degree as with the jet.  Energy-driven feedback depends on the thermal energy the jet deposits at the terminal shock.  Conversely, the wind feedback is strongly dependent on its heavy momentum.    

Our simulations exhibit many features in agreement with past observations and simulation.  The locations of the star forming regions agree well with observations of a radio quiet quasar with dual conical outflows from \citet{Cresci15}, who observe a ring of star formation 2.3 kpc from the center of the galaxy.  This result matches the rings and ovals of star formation as shown in Figures \ref{fig:star_time_mass} and \ref{fig:time_disk_r}, which begin at a radius around 2 kpc and then extend outward with time.  They also observe two clouds at a radius of 1.2 kpc that are forming stars and contend that the compression from the wind has triggered the star formation.  \citet{Tremblay16} sees similar phenomenon but with GMCs being compressed and pushed outward by AGN while forming stars.  We see the same effect in Figure \ref{fig:star_time_mass}, particularly the locations of star formation in wind-i90, which leaves a trail of star formation as clouds are accelerated radially from the nuclear region of the galaxy.  

We see simultaneous positive and negative feedback in all of our simulations as discussed in \citet{Wagner16}.  The negative feedback occurs in the center of the galaxy very quickly, within a radius of 2-3 kpc and within $\sim$3 Myr after the feedback is initialized as shown in Figure \ref{fig:time_disk_r}.  However, the same figure shows positive feedback at radii greater than 3 kpc after 1 Myr.  

The asymmetry of the locations of star forming regions as shown in Figure \ref{fig:star_time_mass} also agrees well with the idea of simultaneous positive and negative feedback as described in \citet{Carniani16}.  They observe two quasars with fast, ionized winds and see negative feedback within the outflow itself but see positive feedback with star formation along the edges of the outflow.  We see a similar result in our simulations, particularly in wind-i45 and wind-i90.  Figure \ref{fig:star_time_mass} shows that far fewer stars are formed within the region of the outflow, to a greater extent as the wind is pointed more directly into the disk in wind-i90.  However, stars are forming around the edges of the outflow.  This phenomenon is highlighted by the quenched central regions without star formation, which have different shapes as a result of the different wind inclinations.  

However, our results seem to conflict with observational indications that jets are more likely to cause positive feedback than radio quiet quasars \citep{Kalf2012}.  \citet{Zinn13} even compares radio-loud and radio quiet galaxies with similar luminosities and find that the galaxies with jets have higher SFR's.  They attribute negative feedback to photo-dissociation and positive feedback to the insertion of mechanical energy.  However, all four of our simulations show positive feedback, with remarkably similar SFRs in three as seen in Figure \ref{fig:sfr}.  

The differences in star formation may exist on smaller scales, where the differing ram pressures of the bubbles will have a strong impact on whether clouds of gas will collapse or be ablated.  \citet{Dugan16} simulates winds of varied ram pressures striking a Bonnor-Ebert sphere of 72 $M_\odot$ and finds an anti-correlation between star formation and wind ram pressure, leading up to a threshold ram pressure above which the wind ablates the cloud before star formation can occur.  Though that threshold ram pressure is above the analytic theoretical expectations shown in Figure \ref{fig:rad_vel_ramp}, we do observe that the ram pressures in bubbles generated by AGN winds are typically an order of magnitude greater than those generated by jets.  That threshold ram pressure of 2e10 dynes cm$^{-2}$ is below the maximum ram pressures observed at the shocks of the bubbles of the jets and winds and above many of the locations inside the bubbles generated from winds.   These ram pressures would indicate that star formation will be more likely in the cocoons from jets than those from winds despite that SFRs in the simulations in this study are similar.

We find an inside out pattern of star formation in gas clouds with large radial velocities in all four simulations.  These results are consistent with those from  \citet{Gaibler12}, \citet{Silk12}, \citet{Ishibashi}, \citet{Dugan14}, \citet{ZubovasKing16} and \citet{Dugan16}.  \citet{Dugan14} evolved the stellar orbits for 1 Gyr after positive feedback from jet simulations and find more random and less coherent stellar velocities with large positive radial and vertical velocities that effectively enlarge the galaxy.  It is reasonable to expect similar patterns of stellar distributions and velocities after a Gyr in the AGN wind simulations performed in this study.

\section{Conclusion} \label{sec:conclusion}

We investigate the differences in mechanical feedback from radio-loud (jet) and radio-quiet (wind) AGN with four hydrodynamic simulations of a single, massive, gas-rich disk galaxy at a redshift of 2--3; one in which the galaxy hosts a jet at an inclination of $0^\circ$ with respect to the galactic plane normal, and three of wide angle, radio-quiet winds with inclinations of $0^\circ$, $45^\circ$, and $90^\circ$.  We analyze the impact of AGN feedback on the host's gas, star formation, and circum-galactic medium.  Jet feedback is energy-driven, while wind feedback is momentum-driven.  Both jets and winds create a pronounced cavity with only little dense gas left in the galactic center where star formation ceases; but we see AGN-triggered star formation at radii greater than $\sim 2$ disk heights or 3 kpc in all the simulations, indicating simultaneous positive and negative feedback in the galaxy at different locations.  Cocoons from jets and winds accelerate clouds of gas where stars are forming, giving these stars larger radial and vertical velocities that may be observable as blue asymmetries in stellar velocity dispersions at different locations.  The jet and winds trigger similar SFRs, but the wind at an inclination of 90$^\circ$ continuously compresses the host's gas, generating high densities the most, thus causing the highest rates of star formation.  More asymmetry in gas distribution and star formation location is created with larger wind inclination.  

Our model generates an expanding ring of triggered star formation with typical velocity of order 1/3 of the circular velocity, superimposed on the older stellar population.  This should result in a potentially detectable blue asymmetry in stellar absorption features at kpc scales (cf. \citet{Cicone16}).

\acknowledgments

ZD was supported by a Centre for Cosmological Studies Balzan Fellowship.  VG was supported by the Sonderforschungsbereich SFB 881 ``The Milky Way System'' (subproject B4) of the German Research Foundation (DFG).  The research of JS has been supported at IAP by  the ERC project  267117 (DARK) hosted by Universit\'e Pierre et Marie Curie - Paris 6   and at JHU by NSF grant OIA-1124403.  We would like to thank Alex Wagner and Rebekka Bieri for their input.  


\end{document}